\documentclass[11pt,a4paper]{article}

\usepackage[utf8]{inputenc}
\usepackage[T1]{fontenc}
\usepackage{amsmath,amssymb,amsfonts}
\usepackage{graphicx}
\usepackage{booktabs}
\usepackage{hyperref}
\usepackage{algorithm}
\usepackage{algpseudocode}
\usepackage{geometry}
\usepackage{xcolor}
\usepackage{CJKutf8}  
\usepackage{tikz}
\usepackage{amsthm}
\usepackage[numbers,sort&compress]{natbib}
\usepackage{multirow}
\usetikzlibrary{3d,calc,arrows.meta}
\theoremstyle{definition}

\newcommand{\sunburst}{\textsc{SunBURST}}
\newcommand{\polychord}{\textsc{PolyChord}}
\newcommand{\multinest}{\textsc{MultiNest}}
\newcommand{\dd}{\mathrm{d}}

\title{\sunburst{}: Deterministic GPU-Accelerated Bayesian Evidence via Mode-Centric Laplace Integration}

\author{
    Ira Wolfson$^{1}$\\[1ex]
    $^1$Department of Electronics and Electrical Engineering\\
    Braude College of Engineering, Israel
}

\date{\today}

\begin{document}

\maketitle

\begin{abstract}
Bayesian evidence evaluation becomes computationally prohibitive in high dimensions due to the curse of dimensionality and the sequential nature of sampling-based methods. We introduce \sunburst{}, a deterministic GPU-native algorithm for Bayesian evidence calculation that replaces global volume exploration with mode-centric geometric integration. The pipeline combines radial mode discovery, batched L-BFGS refinement, and Laplace-based analytic integration, treating modes independently and converting large batches of likelihood evaluations into massively parallel GPU workloads.

For Gaussian and near-Gaussian posteriors, where the Laplace approximation is exact or highly accurate, \sunburst{} achieves numerical agreement at double-precision tolerance in dimensions up to 1024 in our benchmarks, with sub-linear wall-clock scaling across the tested range. In multimodal Gaussian mixtures, conservative configurations yield sub-percent accuracy while maintaining favorable scaling.

\sunburst{} is not intended as a universal replacement for sampling-based inference. Its design targets regimes common in physical parameter estimation and inverse problems, where posterior mass is locally well approximated by Gaussian structure around a finite number of modes. In strongly non-Gaussian settings, the method can serve as a fast geometry-aware evidence estimator or as a preprocessing stage for hybrid workflows. These results show that high-precision Bayesian evidence evaluation can be made computationally tractable in very high dimensions through deterministic integration combined with massive parallelism.

\end{abstract}
\section{Introduction}
\label{sec:intro}

Bayesian model comparison requires computing the marginal likelihood or \emph{evidence}:
\begin{equation}
    \mathcal{Z} = \int \mathcal{L}(\theta) \pi(\theta) \, \dd\theta
\end{equation}
where $\mathcal{L}(\theta) = p(D|\theta)$ is the likelihood and $\pi(\theta)$ is the prior. 
The ratio of evidences between competing models yields the Bayes factor, providing a 
principled framework for model selection that automatically penalizes complexity through 
Occam's razor \citep{mackay2003information, kass1995bayes}. In cosmology, evidence-based 
model comparison has become essential for distinguishing between inflationary scenarios, 
dark energy parameterizations, and extensions to $\Lambda$CDM \citep{trotta2008bayes}.

Computing $\mathcal{Z}$ is notoriously difficult. The integral spans the full prior volume, 
which grows exponentially with dimension $d$, while the posterior mass concentrates in an 
exponentially small fraction of this volume---the well-known \emph{curse of dimensionality} 
\citep{bellman1957dynamic}. Traditional Monte Carlo methods struggle because random sampling 
almost never hits the relevant region; importance sampling requires a proposal distribution 
that already approximates the posterior; and thermodynamic integration demands careful 
temperature scheduling across the full prior-to-posterior transition.

\subsection{Nested Sampling}

Nested sampling \citep{skilling2004nested, skilling2006nested} elegantly transforms the 
multi-dimensional evidence integral into a one-dimensional integral over the prior mass $X$:
\begin{equation}
    \mathcal{Z} = \int_0^1 \mathcal{L}(X) \, \dd X
\end{equation}
where $X(\lambda) = \int_{\mathcal{L}(\theta) > \lambda} \pi(\theta) \, \dd\theta$ is the 
prior volume enclosed by the iso-likelihood contour $\mathcal{L} = \lambda$. This 
transformation converts the problem from sampling a complex posterior to iteratively 
compressing the prior volume while tracking the likelihood threshold.

Modern implementations have achieved remarkable success in astrophysical inference. 
\textsc{MultiNest} \citep{feroz2008multinest, feroz2009multinest} introduced ellipsoidal 
decomposition for multimodal posteriors. \textsc{PolyChord} \citep{handley2015polychord} 
replaced rejection sampling with slice sampling, improving efficiency up to $d \sim 50$. 
More recent implementations include \textsc{dynesty} \citep{speagle2020dynesty} with dynamic 
live-point allocation, \textsc{UltraNest} \citep{buchner2021ultranest} emphasizing robustness, 
and \textsc{JAXNS} \citep{albert2023jaxns} leveraging JAX for GPU acceleration.

Despite these advances, nested sampling faces fundamental scaling limitations. The 
computational cost scales as $O(n_\text{live} \cdot n_\text{iter})$, where 
$n_\text{iter} \propto n_\text{live} \cdot H$ for posterior information $H$, 
and $n_\text{live} = O(K \cdot d)$ for $K$ modes in $d$ dimensions 
\citep{handley2015polychord}. This yields $O(K \cdot d^2)$ scaling that becomes 
prohibitive beyond $d \approx 50$--$100$.

\subsection{The GPU Opportunity}

Graphics processing units offer massive parallelism, yet nested sampling's sequential 
contraction of live points prevents straightforward acceleration. Each iteration depends 
on the previous likelihood threshold, limiting GPU speedup to parallelizing individual 
likelihood evaluations.

For posteriors that are well-approximated by mixtures of Gaussians, we observe that 
the evidence integral admits an entirely different computational strategy:
\begin{enumerate}
    \item Detect posterior modes via parallel optimization
    \item Refine mode locations and estimate local curvature
    \item Compute per-mode evidence via the Laplace approximation \citep{tierney1986accurate}
    \item Sum contributions across modes
\end{enumerate}
Each step is embarrassingly parallel, enabling full GPU utilization.

\subsection{Contributions}

We introduce \sunburst{} (\textbf{S}eeded \textbf{U}niverse \textbf{N}avigation---\textbf{B}ayesian 
\textbf{U}nification via \textbf{R}adial \textbf{S}hooting \textbf{T}echniques), a GPU-accelerated 
evidence calculator that achieves:
\begin{enumerate}
    \item \textbf{$O(K)$ mode scaling}: Evidence is computed per-mode, requiring only $O(K)$ 
          mode characterizations rather than $O(K \cdot d)$ live points
    \item \textbf{Sub-linear wall-clock scaling}: Overall $O(d^{0.7})$ scaling from 2D 
      to 1024D, with absolute times under 7 seconds at 1024D on consumer GPU (for Gaussian Benchmark)
    \item \textbf{Automatic mode detection}: Batched optimization discovers posterior modes 
          without prior knowledge of their number or locations
    \item \textbf{Numerical precision}: relative error below $10^{-12}$ for Gaussian posteriors through 1024D

\end{enumerate}

The pipeline comprises three core modules, named after Guang Ping Yang Style Tai Chi forms 
in honor of Master Donald Rubbo:
\begin{itemize}
    \item \textbf{CarryTiger} (\begin{CJK}{UTF8}{gbsn}抱虎归山\end{CJK}, 
          \emph{B\`ao H\v{u} Gu\=i Sh\=an}---``Carry Tiger to Mountain''): 
          Mode detection via multi-scale ray casting and batched L-BFGS optimization 
          with convergence-anticonvergence oscillations
    \item \textbf{GreenDragon} (\begin{CJK}{UTF8}{gbsn}青龙出水\end{CJK}, 
          \emph{Q\=ing L\'ong Ch\=u Shu\v{i}}---``Green Dragon Rises from Water''): 
          Peak refinement using GPU-parallel L-BFGS \citep{liu1989lbfgs} with 
          Hessian estimation and saddle-point filtering
    \item \textbf{BendTheBow} (\begin{CJK}{UTF8}{gbsn}弯弓射虎\end{CJK}, 
          \emph{W\=an G\=ong Sh\`e H\v{u}}---``Bend the Bow, Shoot the Tiger''): 
          Evidence calculation via Laplace approximation with adaptive Hessian 
          estimation for general Gaussian geometries
\end{itemize}

\sunburst{} is not intended as a universal replacement for sampling-based evidence estimators.
Its design targets the regime in which posterior mass is well approximated locally by Gaussian structure around a finite number of modes — a regime common in physical parameter estimation, model calibration, and inverse problems. In strongly non-Gaussian settings (e.g., pronounced curvature, phase transitions, or heavy-tailed structure), \sunburst{} should instead be viewed as a fast geometry-aware evidence estimator or as a preprocessing stage within hybrid inference workflows.

To support such comprehensive approaches, an optional fourth module, \textbf{GraspBirdsTail} (\begin{CJK}{UTF8}{gbsn}揽雀尾\end{CJK}), performs eigendecomposition-based dimensional reduction for handoff to downstream samplers when full posterior characterization is required.

\sunburst{} is implemented in Python using CuPy \citep{cupy2017} for GPU acceleration, 
with automatic fallback to NumPy \citep{harris2020numpy} and SciPy \citep{virtanen2020scipy} 
for CPU-only systems.


\section{The Curse of Dimensionality}
\label{sec:curse}

Before presenting our algorithm, we must confront the fundamental obstacle that makes high-dimensional Bayesian evidence calculation so difficult. The ``curse of dimensionality''---a term coined by \citet{bellman1961adaptive}---refers to a collection of pathologies that emerge as the number of dimensions grows. For evidence calculation, these pathologies conspire to make the problem exponentially harder in ways that are not merely inconvenient but genuinely devastating.

\textbf{Readers familiar with these pathologies may skip to Section~\ref{sec:curse-summary} for a summary of the key challenges that \sunburst{} must overcome.}

\subsection{The Concentration of Measure}

Consider a $d$-dimensional hypercube $[-1,1]^d$ with volume $V_d = 2^d$. The volume of the inscribed hypersphere of radius 1 is:
\begin{equation}
    V_{\text{sphere}}(d) = \frac{\pi^{d/2}}{\Gamma(d/2 + 1)}
\end{equation}
where $\Gamma(z)$ is the gamma function, the continuous extension of the factorial: $\Gamma(n) = (n-1)!$ for positive integers, with $\Gamma(1/2) = \sqrt{\pi}$. The key property driving the curse is that $\Gamma(d/2 + 1)$ grows superexponentially---faster than $\pi^{d/2}$---causing the sphere volume to collapse.

The ratio $V_{\text{sphere}}/V_{\text{cube}}$ reveals the first horror:

\begin{center}
\begin{tabular}{r|cccccc}
$d$ & 2 & 10 & 50 & 100 & 500 & 1000 \\
\hline
$V_{\text{sphere}}/V_{\text{cube}}$ & 0.79 & $2.5 \times 10^{-3}$ & $1.9 \times 10^{-24}$ & $1.9 \times 10^{-70}$ & $\sim 10^{-616}$ & $\sim 10^{-1464}$
\end{tabular}
\end{center}

By 100 dimensions, the sphere occupies a fraction of the cube smaller than the ratio of a proton's volume to the observable universe. By 500 dimensions, we have exhausted the dynamic range of double-precision floating point ($10^{-308}$). The geometry of high-dimensional space is dominated by corners.

\subsection{The Shell Theorem}

Where does the mass of a high-dimensional Gaussian live? For $\mathcal{N}(0, I_d)$, the radial distribution of probability mass is:
\begin{equation}
    p(r) \propto r^{d-1} e^{-r^2/2}
\end{equation}

This peaks at $r^* = \sqrt{d-1} \approx \sqrt{d}$, with standard deviation $\sigma_r \approx 1/\sqrt{2}$ \emph{independent of $d$}. The probability mass concentrates in a thin shell:

\begin{center}
\begin{tabular}{r|ccc}
$d$ & Shell radius $r^*$ & Shell thickness $\sigma_r$ & Relative thickness \\
\hline
10 & 3.0 & 0.71 & 24\% \\
100 & 10.0 & 0.71 & 7.1\% \\
1000 & 31.6 & 0.71 & 2.2\% \\
\end{tabular}
\end{center}

The mode of a high-dimensional Gaussian (at the origin) contains \emph{negligible} probability mass. Almost all the probability lives in a shell of radius $\sqrt{d}$ and thickness $O(1)$---a set of measure zero as $d \to \infty$ relative to the enclosing ball.

\subsection{The Distance Catastrophe}

The Euclidean distance between two points $x, y \in \mathbb{R}^d$ is:
\begin{equation}
    d(x, y) = \sqrt{\sum_{i=1}^{d} (x_i - y_i)^2}
\end{equation}

Consider two random points uniformly distributed in the unit hypercube $[0,1]^d$. Each term $(x_i - y_i)^2$ is an independent random variable with mean $\mu = \mathbb{E}[(x_i - y_i)^2] = 1/6$ and finite variance. The sum of $d$ such terms has mean $d/6$, so:
\begin{equation}
    \mathbb{E}[d(x,y)] \approx \sqrt{d/6} \xrightarrow{d \to \infty} \infty
\end{equation}

The expected distance between \emph{any} two random points diverges. But the standard deviation of the sum grows only as $\sqrt{d}$, so the coefficient of variation (standard deviation divided by mean) vanishes:
\begin{equation}
    \frac{\text{Std}[d(x,y)]}{\mathbb{E}[d(x,y)]} \sim \frac{\sqrt{d}}{d} = \frac{1}{\sqrt{d}} \to 0
\end{equation}

The distance to your nearest neighbor and the distance to your farthest neighbor both concentrate around $\sqrt{d/6}$. In the limit, \emph{all points are equidistant from all other points}. The concepts of ``nearby'' and ``far away'' lose meaning. Clustering algorithms fail. Nearest-neighbor search becomes arbitrary. The geometry we rely on in 2D and 3D intuition simply does not exist.

\paragraph{The $L_\infty$ Metric.} The $L_\infty$ metric does not suffer from the divergence 
pathology of $L_2$:
\begin{equation}
    d_\infty(x, y) = \max_{i=1,\ldots,d} |x_i - y_i|
\end{equation}
Since this is a maximum rather than a sum, it remains bounded: for points in $[0,1]^d$, 
we have $d_\infty(x,y) \in [0,1]$ regardless of dimension. However, recent work shows 
that $L_\infty$ distances between \emph{random} points still concentrate toward a common 
value as $d \to \infty$, rendering nearest-neighbor search unreliable \citep{peng2024curse}. 
For \sunburst{}'s use cases---deduplicating peaks that have already converged to similar 
locations, and detecting convergence relative to a known reference---$L_\infty$ remains 
effective because we compare structured points from optimization trajectories, not 
uniformly random samples.

\subsection{The Sampling Catastrophe}

To uniformly cover a $d$-dimensional unit hypercube with grid spacing $\delta$, we need $N = (1/\delta)^d$ points. For $\delta = 0.1$ (10 points per dimension):

\begin{center}
\begin{tabular}{r|ccc}
$d$ & Grid points & At $10^9$/sec & Storage (64-bit) \\
\hline
10 & $10^{10}$ & 10 seconds & 80 GB \\
20 & $10^{20}$ & 3,000 years & $8 \times 10^{11}$ TB \\
50 & $10^{50}$ & $3 \times 10^{33}$ years & --- \\
100 & $10^{100}$ & $3 \times 10^{83}$ years & --- \\
\end{tabular}
\end{center}

A grid with spacing 0.1 in 100 dimensions requires more points than there are atoms in the observable universe ($\sim 10^{80}$). Monte Carlo methods fare no better: to achieve relative error $\epsilon$ on an integral via naive sampling requires $N = O(\epsilon^{-2})$ samples \emph{that actually hit the relevant region}. When the posterior occupies a fraction $f \sim 10^{-70}$ of the prior volume, we need $N \sim 10^{70}$ samples before a single one lands in the posterior.

\subsection{The Evidence Calculation Catastrophe}

For Bayesian evidence, these pathologies compound. The evidence integral:
\begin{equation}
    \mathcal{Z} = \int_{\Theta} \mathcal{L}(\theta) \pi(\theta) \, d\theta
\end{equation}
requires integrating over the full prior volume $V_\Theta$, but the posterior mass concentrates in an exponentially small region. The \emph{compression ratio}---the ratio of prior to posterior volume---grows exponentially:
\begin{equation}
    \frac{V_{\text{prior}}}{V_{\text{posterior}}} \sim \left(\frac{\sigma_{\text{prior}}}{\sigma_{\text{posterior}}}\right)^d
\end{equation}

For a prior 10$\times$ wider than the posterior in each dimension, this ratio is $10^d$. At $d = 100$, we must accurately integrate over a region occupying $10^{-100}$ of the prior volume.

Nested sampling addresses this through the prior-mass transformation, but pays a different price: the number of iterations scales as $n_{\text{iter}} \sim n_{\text{live}} \cdot H$, where the information $H = \int p(\theta|D) \log[p(\theta|D)/\pi(\theta)] \, d\theta$ typically grows as $O(d)$. Combined with $n_{\text{live}} = O(K \cdot d)$ for $K$ modes, the total cost becomes $O(K \cdot d^2)$---still polynomial, but steep.

\subsection{The Multimodal Catastrophe}

Multiple posterior modes transform difficulty into impossibility. If $K$ modes are separated by low-likelihood valleys, nested sampling must maintain enough live points to populate each mode. The required $n_{\text{live}} = O(K \cdot d)$ ensures at least one live point per mode, but at high $d$ this becomes prohibitive.

Worse, mode detection itself becomes unreliable. Random sampling in high dimensions almost never discovers modes---the probability of landing within distance $\epsilon$ of any given point scales as $\epsilon^d$. Without prior knowledge of mode locations, algorithms resort to expensive exploration phases.

\subsection{Summary: The Five Facets of the Curse}
\label{sec:curse-summary}

The curse of dimensionality manifests through five distinct but interrelated pathologies that any high-dimensional evidence calculator must overcome:

\begin{enumerate}
    \item \textbf{Volume collapse}: The ratio of hypersphere to hypercube volume vanishes superexponentially---by $d=100$, the sphere occupies $<10^{-70}$ of the cube, exhausting floating-point precision by $d \approx 500$.
    
    \item \textbf{Shell concentration}: Probability mass migrates from modes to shells. A $d$-dimensional Gaussian concentrates in a shell of radius $\sqrt{d}$ and thickness $O(1)$; the mode contains negligible mass.
    
    \item \textbf{Distance concentration}: Under $L_2$, all pairwise distances converge to $\sqrt{d/6}$---nearest and farthest neighbors become indistinguishable. $L_\infty$ avoids divergence but still concentrates for random points.
    
    \item \textbf{Sampling explosion}: Grid coverage requires $(1/\delta)^d$ points; Monte Carlo requires $O(f^{-1})$ samples where $f \sim 10^{-d}$ is the posterior-to-prior volume ratio.
    
    \item \textbf{Integration catastrophe}: Evidence calculation must integrate over a posterior occupying $\sim 10^{-d}$ of the prior volume. Nested sampling costs $O(K \cdot d^2)$; naive methods are exponential.
\end{enumerate}

Each of these pathologies motivates a corresponding design constraint that is addressed explicitly in the \sunburst{} algorithm described in Section~\ref{sec:algorithm}.

\subsection{The Geometry of Escape}

How can any algorithm escape these exponential barriers? The curse is not merely a computational inconvenience---it reflects genuine mathematical structure. No algorithm can uniformly sample high-dimensional spaces efficiently.

The escape routes are:
\begin{enumerate}
    \item \textbf{Exploit structure}: If the posterior has special structure (e.g., near-Gaussian, separable, sparse), exploit it ruthlessly
    \item \textbf{Work locally}: Abandon global coverage; focus computational resources on the posterior mass
    \item \textbf{Parallelize}: Trade sequential depth for parallel breadth via GPU
    \item \textbf{Reuse samples}: Never discard likelihood evaluations; bank them for later stages
\end{enumerate}

\sunburst{} employs all four strategies. The following section describes how.


%
%

\section{Algorithm}
\label{sec:algorithm}

\subsection{Overview}

\sunburst{} computes Bayesian evidence through a preprocessing step followed by three main modules:

\begin{itemize}
    \item \textbf{Preprocessing step}: Detects and removes pathological dimensions (flat, soft nuisance) via $O(2d)$ sensitivity probing
    \item \textbf{Module 1 (CarryTiger)}: Discovers posterior modes via ray casting through the prior hypercube, followed by ChiSao exploration with convergence-anticonvergence oscillations
    \item \textbf{Module 2 (GreenDragon)}: Refines peak locations to within $<10^{-12}$ error via batched L-BFGS, producing the TrajectoryBank
    \item \textbf{Module 3 (BendTheBow)}: Computes per-mode evidence via Laplace approximation, using the TrajectoryBank for zero-cost rotation detection
\end{itemize}

An optional fourth module (\textbf{GraspBirdsTail}) performs eigendecomposition-based dimensional reduction for handoff to downstream samplers.

Sample banks persist all likelihood evaluations in GPU memory, enabling geometry detection without redundant computation.

\subsubsection{Preprocessing}

Before mode detection, we probe the likelihood to detect pathological dimensions:
\begin{itemize}
    \item \textbf{Flat dimensions}: Zero sensitivity---the likelihood ignores these parameters entirely
    \item \textbf{Soft nuisance dimensions}: Very weak constraints that would create conditioning problems
\end{itemize}

Pathological dimensions are marginalized analytically, and the remaining ``active'' dimensions proceed to Module 1.

\subsubsection{Module 1: CarryTiger}

We cast rays through the prior hypercube using four complementary geometries (vertex-to-vertex, vertex-to-edge, wall-to-wall, and sunburst\footnote{A form of ray casting centered at some interior point, analogous to rays emanating from a sun. Hence the name.} from center). Along each ray, we sample in two passes:

\begin{enumerate}
    \item \textbf{Coarse pass}: Uniform sampling to discover where the likelihood is non-negligible
    \item \textbf{Refinement pass}: Adaptive resampling concentrated where the coarse pass found significant likelihood
\end{enumerate}

Starting from promising ray samples, ChiSao exploration locates posterior modes through convergence-anticonvergence oscillations. Samples that reach true peaks ``stick'' and remain frozen while others continue exploring.

\subsubsection{Module 2: GreenDragon}

Coarse peak locations from Module 1 are refined to within $10^{-12}$ error via:
\begin{enumerate}
    \item Seeding multiple starting points around each coarse peak
    \item Batched L-BFGS with tight convergence tolerances
    \item Deduplication and saddle-point filtering
\end{enumerate}

Precise peak locations are crucial: the Laplace approximation error scales with distance from the true maximum. L-BFGS trajectories are stored in the TrajectoryBank for use in Module 3.

\subsubsection{Module 3: BendTheBow}

For each refined peak, we compute the evidence contribution via Laplace approximation:
\begin{enumerate}
    \item \textbf{Geometry detection}: Check if posterior has off-diagonal correlations (using GreenDragon's TrajectoryBank or finite-difference probing)
    \item \textbf{Hessian computation}: Diagonal-only for axis-aligned, full matrix for rotated posteriors
    \item \textbf{Laplace approximation}: Analytic Gaussian integral using the Hessian determinant
\end{enumerate}

The total evidence is the sum (logsumexp) over all modes.

\subsection{Problem Statement}

Given a log-likelihood function $\ell(\theta) = \log \mathcal{L}(\theta)$ and uniform prior $\pi(\theta)$ over a $d$-dimensional hypercube $\Theta = \prod_{i=1}^d [a_i, b_i]$, we seek the log-evidence:
\begin{equation}
    \log Z = \log \int_\Theta \exp(\ell(\theta)) \, d\theta - \log V_\Theta
\end{equation}

\paragraph{Prior Assumption.} The current implementation assumes the likelihood is well-contained within the prior hypercube---that is, the prior boundaries do not truncate any significant likelihood mass. Under this assumption, the prior volume $V_\Theta = \prod_i (b_i - a_i)$ enters only as a normalizing constant. Handling informative or non-uniform priors that interact non-trivially with the likelihood is deferred to future work.

\subsection{Module Structure}

\begin{center}
\begin{tabular}{l|l|l}
\textbf{Module} & \textbf{Name} & \textbf{Output} \\
\hline
- & Preprocessing & Active dimensions, marginal corrections \\
1 & CarryTiger & Coarse peaks + RayBank \\
2 & GreenDragon & Refined peaks + Hessians + TrajectoryBank \\
3 & BendTheBow & $\log Z$ \\
4 & GraspBirdsTail & Reduced likelihood (optional) \\
\end{tabular}
\end{center}

\textbf{Sample Banks} persist all likelihood evaluations in GPU memory:
\begin{itemize}
    \item \textbf{RayBank}: Ray geometry, sample positions, log-likelihoods from Module 1 (CarryTiger)
    \item \textbf{TrajectoryBank}: L-BFGS trajectories from Module 2 (GreenDragon), used for rotation detection in Module 3
\end{itemize}

These banks enable geometry detection in Module 3 without additional likelihood evaluations.

\subsection{Preprocessing: Dimensional Pathology Detection}
\label{sec:precheck}

Before mode detection, \sunburst{} performs a rapid $O(2d)$ probe to detect pathological dimensions that would cause numerical difficulties in downstream stages.

\subsubsection{Sensitivity Probing}

For each dimension $i$, we evaluate the likelihood sensitivity by perturbing from the prior center to the boundaries:
\begin{equation}
    s_i = |\ell(\theta_{\text{center}} + \Delta_i e_i) - \ell(\theta_{\text{center}})| + 
          |\ell(\theta_{\text{center}} - \Delta_i e_i) - \ell(\theta_{\text{center}})|
\end{equation}
where $\Delta_i = (b_i - a_i)/2$ is the half-width of the prior in dimension $i$, and $e_i$ is the $i$-th unit vector. This requires exactly $2d$ likelihood evaluations.

\subsubsection{Flat Dimensions}

Dimensions where the likelihood is completely insensitive:
\begin{equation}
    s_i < \epsilon_{\text{flat}} \cdot \text{median}(s_1, \ldots, s_d)
\end{equation}
with default $\epsilon_{\text{flat}} = 10^{-6}$. These represent parameters that the data cannot constrain at all---the posterior equals the prior.

\subsubsection{Soft Nuisance Dimensions}

Dimensions with very weak constraints that would create conditioning problems:
\begin{equation}
    s_i < \epsilon_{\text{soft}} \cdot \max(s_1, \ldots, s_d) \quad \text{and} \quad s_i \geq \epsilon_{\text{flat}} \cdot \text{median}(s)
\end{equation}
with default $\epsilon_{\text{soft}} = 10^{-3}$. This threshold catches condition numbers $\kappa > 30$.

\subsubsection{Analytical Marginalization}

Flagged dimensions are removed from the active problem and marginalized analytically:

\paragraph{Flat dimensions} contribute $\log(\Delta_i)$ to the evidence (uniform prior integral).

\paragraph{Soft nuisance dimensions} contribute a Gaussian integral. We estimate the effective width $\sigma_i$ from the sensitivity:
\begin{equation}
    \sigma_i \approx \frac{\Delta_i}{\sqrt{s_i}}, \quad \text{contribution} = \frac{1}{2}\log(2\pi) + \log(\sigma_i)
\end{equation}\\

\paragraph{Assumption underlying nuisance-dimension marginalization.}
~The Gaussian marginal correction applied to ``soft nuisance'' dimensions assumes that, along each such coordinate, the log-likelihood is locally well approximated by a one-dimensional quadratic around the probed reference point (the prior center). Under this assumption, the log-likelihood drop measured between the center and the prior bounds provides a proxy for local curvature, allowing estimation of an effective Gaussian width $\sigma_i$. The resulting contribution $\tfrac12\log(2\pi)+\log\sigma_i$ corresponds to analytically marginalizing a locally Gaussian factor in that coordinate.

When the likelihood is not approximately quadratic on this scale, or when the prior center lies far from the local maximum in that coordinate, this correction should be interpreted as a heuristic scale estimate rather than an exact marginalization. In such cases, \sunburst{}'s main Laplace stage (which uses curvature at detected modes) remains the primary source of evidence accuracy.

The remaining $d_{\text{active}}$ dimensions proceed to CarryTiger. This preprocessing prevents catastrophic condition numbers in downstream Hessian computations.

\subsubsection{Algorithm Summary}

\begin{algorithm}[H]
\caption{Preprocessing: Dimensional Pathology Detection}
\begin{algorithmic}[1]
\Require Log-likelihood $\ell$, bounds $[a_i, b_i]$, thresholds $\epsilon_{\text{flat}}, \epsilon_{\text{soft}}$
\Ensure Active dimensions, marginal corrections

\State $\theta_{\text{center}} \gets (a + b) / 2$
\State $\ell_0 \gets \ell(\theta_{\text{center}})$
\For{$i = 1, \ldots, d$} \Comment{GPU-parallel: all $2d$ evals in one batch}
    \State $\Delta_i \gets (b_i - a_i) / 2$
    \State $s_i \gets |\ell(\theta_{\text{center}} + \Delta_i e_i) - \ell_0| + |\ell(\theta_{\text{center}} - \Delta_i e_i) - \ell_0|$
\EndFor
\State $\texttt{flat} \gets \{i : s_i < \epsilon_{\text{flat}} \cdot \text{median}(s)\}$
\State $\texttt{soft} \gets \{i : s_i < \epsilon_{\text{soft}} \cdot \max(s)\} \setminus \texttt{flat}$
\State $\texttt{active} \gets \{1, \ldots, d\} \setminus (\texttt{flat} \cup \texttt{soft})$
\State $\log Z_{\text{marginal}} \gets \sum_{i \in \texttt{flat}} \log(\Delta_i) + \sum_{i \in \texttt{soft}} \left[\frac{1}{2}\log(2\pi) + \log(\Delta_i / \sqrt{s_i})\right]$
\State \Return $\texttt{active}, \log Z_{\text{marginal}}$
\end{algorithmic}
\end{algorithm}

\subsection{Module 1: CarryTiger (\begin{CJK}{UTF8}{bsmi}抱虎歸山\end{CJK} -- Carry Tiger to Mountain)}
\label{sec:CarryTiger}

CarryTiger discovers posterior modes through ray casting and ChiSao exploration.

\subsubsection{Ray Generation}

Four ray types provide complementary coverage of the prior hypercube (Figure~\ref{fig:raycasting}):
\begin{itemize}
    \item \textbf{Vertex-to-Vertex (V2V)}: Connect random vertex pairs---these rays traverse the full diagonal extent
    \item \textbf{Vertex-to-Edge (V2E)}: From vertices toward edge midpoints---intermediate angles
    \item \textbf{Wall-to-Wall (W2W)}: Between random points on opposing faces---guaranteed to span the full extent in one coordinate direction
    \item \textbf{Sunburst}: From hypercube center in uniformly distributed directions on $S^{d-1}$---isotropic coverage
\end{itemize}

The number of rays adapts to dimension: $n_\text{rays} = 10 + \log_2(d)$.

\begin{figure}[htbp]
\centering
\resizebox{\textwidth}{!}{%
\begin{tikzpicture}[
    box2d/.style={thick, black!70},
    box3d/.style={thick, black!70},
    box3d back/.style={thick, black!35, dashed},
    ray/.style={thick, -{Stealth[length=2.5mm]}, #1},
    ray outside/.style={thick, #1, densely dotted},
    sample coarse/.style={circle, draw=#1, fill=white, inner sep=1.3pt, line width=0.4pt},
    sample refined/.style={circle, fill=#1, inner sep=1.5pt},
]

\def\panelW{5.5}
\def\gap{0.3}
\def\L{2.6}

\def\colA{0}
\def\colB{\panelW + \gap}
\def\colC{2*\panelW + 2*\gap}
\def\colD{3*\panelW + 3*\gap}

\def\rowTop{0}
\def\rowBot{-9.5}


\begin{scope}[shift={(\colA + 0.8, \rowTop)},
    x={(-0.65cm,-0.95cm)}, y={(1.6cm,0cm)}]
    
    \def\S{2.6}
    \draw[box2d] (0,0) -- (\S,0) -- (\S,\S) -- (0,\S) -- cycle;
    
    \def\cx{1.3}
    \def\cy{1.3}
    \draw[red!80!black, thick] (\cx,\cy) ellipse (0.14 and 0.12);
    \draw[red!60!black, thick] (\cx,\cy) ellipse (0.30 and 0.25);
    \draw[red!40!black, thick] (\cx,\cy) ellipse (0.48 and 0.40);
    \draw[red!25!black, thick] (\cx,\cy) ellipse (0.70 and 0.58);
    \node[red!60!black, font=\small] at (\cx+0.18, \cy+0.78) {$\mathcal{L}$};
    
    \draw[ray outside=blue!70!black] (-0.2, -0.2) -- (0,0);
    \draw[ray=blue!70!black] (0,0) -- (\S+0.2, \S+0.2);
    \foreach \t in {0.15, 0.4, 0.65, 0.9} {
        \node[sample coarse=blue!70!black] at (\t*\S, \t*\S) {};
    }
    \foreach \t in {0.45, 0.55} {
        \node[sample refined=blue!70!black] at (\t*\S, \t*\S) {};
    }
\end{scope}

\node[font=\bfseries\Large] at (\colA + 2.0, \rowBot - 1.8) {V2V};

\begin{scope}[shift={(\colA + 0.8, \rowBot)},
    x={(-0.65cm,-0.48cm)}, y={(1.65cm,0cm)}, z={(0cm,1.65cm)}]
    
    \draw[box3d back] (0,0,0) -- (\L,0,0);
    \draw[box3d back] (0,0,0) -- (0,\L,0);
    \draw[box3d back] (0,0,0) -- (0,0,\L);
    \draw[box3d back] (\L,0,0) -- (\L,\L,0);
    \draw[box3d back] (0,\L,0) -- (\L,\L,0);
    \draw[box3d back] (\L,0,0) -- (\L,0,\L);
    \draw[box3d back] (0,\L,0) -- (0,\L,\L);
    \draw[box3d back] (0,0,\L) -- (\L,0,\L);
    \draw[box3d back] (0,0,\L) -- (0,\L,\L);
    
    \shade[ball color=red!60!black, opacity=0.6] (1.3,1.3,1.3) circle (0.38);
    
    \draw[box3d] (\L,\L,0) -- (\L,\L,\L);
    \draw[box3d] (\L,0,\L) -- (\L,\L,\L);
    \draw[box3d] (0,\L,\L) -- (\L,\L,\L);
    
    \draw[ray outside=blue!70!black] (-0.2,-0.2,-0.2) -- (0,0,0);
    \draw[ray=blue!70!black] (0,0,0) -- (\L+0.2,\L+0.2,\L+0.2);
    \foreach \t in {0.15, 0.4, 0.65, 0.9} {
        \node[sample coarse=blue!70!black] at (\t*\L, \t*\L, \t*\L) {};
    }
    \foreach \t in {0.45, 0.55} {
        \node[sample refined=blue!70!black] at (\t*\L, \t*\L, \t*\L) {};
    }
\end{scope}


\begin{scope}[shift={(\colB + 0.8, \rowTop)},
    x={(-0.65cm,-0.95cm)}, y={(1.6cm,0cm)}]
    
    \def\S{2.6}
    \draw[box2d] (0,0) -- (\S,0) -- (\S,\S) -- (0,\S) -- cycle;
    
    \def\cx{1.3}
    \def\cy{1.3}
    \draw[red!80!black, thick] (\cx,\cy) ellipse (0.14 and 0.12);
    \draw[red!60!black, thick] (\cx,\cy) ellipse (0.30 and 0.25);
    \draw[red!40!black, thick] (\cx,\cy) ellipse (0.48 and 0.40);
    \draw[red!25!black, thick] (\cx,\cy) ellipse (0.70 and 0.58);
    \node[red!60!black, font=\small] at (\cx+0.18, \cy+0.78) {$\mathcal{L}$};
    
    \draw[ray outside=green!55!black] (-0.2, 0.5) -- (0, 0.5);
    \draw[ray=green!55!black] (0, 0.5) -- (\S, 1.6);
    \draw[ray outside=green!55!black] (\S, 1.6) -- (\S+0.2, 1.72);
    \foreach \t in {0.15, 0.4, 0.65, 0.9} {
        \node[sample coarse=green!55!black] at (\t*\S, {0.5 + \t*1.1}) {};
    }
    \foreach \t in {0.45, 0.58} {
        \node[sample refined=green!55!black] at (\t*\S, {0.5 + \t*1.1}) {};
    }
\end{scope}

\node[font=\bfseries\Large] at (\colB + 2.0, \rowBot - 1.8) {W2W};

\begin{scope}[shift={(\colB + 0.8, \rowBot)},
    x={(-0.65cm,-0.48cm)}, y={(1.65cm,0cm)}, z={(0cm,1.65cm)}]
    
    \draw[box3d back] (0,0,0) -- (\L,0,0);
    \draw[box3d back] (0,0,0) -- (0,\L,0);
    \draw[box3d back] (0,0,0) -- (0,0,\L);
    \draw[box3d back] (\L,0,0) -- (\L,\L,0);
    \draw[box3d back] (0,\L,0) -- (\L,\L,0);
    \draw[box3d back] (\L,0,0) -- (\L,0,\L);
    \draw[box3d back] (0,\L,0) -- (0,\L,\L);
    \draw[box3d back] (0,0,\L) -- (\L,0,\L);
    \draw[box3d back] (0,0,\L) -- (0,\L,\L);
    
    \shade[ball color=red!60!black, opacity=0.6] (1.3,1.3,1.3) circle (0.38);
    
    \draw[box3d] (\L,\L,0) -- (\L,\L,\L);
    \draw[box3d] (\L,0,\L) -- (\L,\L,\L);
    \draw[box3d] (0,\L,\L) -- (\L,\L,\L);
    
    \draw[ray outside=green!55!black] (-0.2,0.5,1.2) -- (0,0.5,1.2);
    \draw[ray=green!55!black] (0,0.5,1.2) -- (\L,1.6,1.1);
    \draw[ray outside=green!55!black] (\L,1.6,1.1) -- (\L+0.2,1.72,1.08);
    \foreach \t in {0.15, 0.4, 0.65, 0.9} {
        \node[sample coarse=green!55!black] at (\t*\L, {0.5 + \t*1.1}, {1.2 - \t*0.1}) {};
    }
    \foreach \t in {0.45, 0.58} {
        \node[sample refined=green!55!black] at (\t*\L, {0.5 + \t*1.1}, {1.2 - \t*0.1}) {};
    }
\end{scope}


\begin{scope}[shift={(\colC + 0.8, \rowTop)},
    x={(-0.65cm,-0.95cm)}, y={(1.6cm,0cm)}]
    
    \def\S{2.6}
    \draw[box2d] (0,0) -- (\S,0) -- (\S,\S) -- (0,\S) -- cycle;
    
    \def\cx{1.3}
    \def\cy{1.3}
    \draw[red!80!black, thick] (\cx,\cy) ellipse (0.14 and 0.12);
    \draw[red!60!black, thick] (\cx,\cy) ellipse (0.30 and 0.25);
    \draw[red!40!black, thick] (\cx,\cy) ellipse (0.48 and 0.40);
    \draw[red!25!black, thick] (\cx,\cy) ellipse (0.70 and 0.58);
    \node[red!60!black, font=\small] at (\cx+0.18, \cy+0.78) {$\mathcal{L}$};
    
    \draw[ray outside=purple!70] (\S+0.2, -0.18) -- (\S, 0);
    \draw[ray=purple!70] (\S, 0) -- (0, \S/2);
    \draw[ray outside=purple!70] (0, \S/2) -- (-0.18, \S/2 + 0.12);
    \foreach \t in {0.12, 0.38, 0.62, 0.88} {
        \node[sample coarse=purple!70] at ({\S*(1-\t)}, {\t*\S/2}) {};
    }
    \foreach \t in {0.35, 0.48} {
        \node[sample refined=purple!70] at ({\S*(1-\t)}, {\t*\S/2}) {};
    }
\end{scope}

\node[font=\bfseries\Large] at (\colC + 2.0, \rowBot - 1.8) {V2E};

\begin{scope}[shift={(\colC + 0.8, \rowBot)},
    x={(-0.65cm,-0.48cm)}, y={(1.65cm,0cm)}, z={(0cm,1.65cm)}]
    
    \draw[box3d back] (0,0,0) -- (\L,0,0);
    \draw[box3d back] (0,0,0) -- (0,\L,0);
    \draw[box3d back] (0,0,0) -- (0,0,\L);
    \draw[box3d back] (\L,0,0) -- (\L,\L,0);
    \draw[box3d back] (0,\L,0) -- (\L,\L,0);
    \draw[box3d back] (\L,0,0) -- (\L,0,\L);
    \draw[box3d back] (0,\L,0) -- (0,\L,\L);
    \draw[box3d back] (0,0,\L) -- (\L,0,\L);
    \draw[box3d back] (0,0,\L) -- (0,\L,\L);
    
    \shade[ball color=red!60!black, opacity=0.6] (1.3,1.3,1.3) circle (0.38);
    
    \draw[box3d] (\L,\L,0) -- (\L,\L,\L);
    \draw[box3d] (\L,0,\L) -- (\L,\L,\L);
    \draw[box3d] (0,\L,\L) -- (\L,\L,\L);
    
    \draw[ray outside=purple!70] (\L+0.2,-0.18,-0.12) -- (\L,0,0);
    \draw[ray=purple!70] (\L,0,0) -- (0,\L/2,\L);
    \draw[ray outside=purple!70] (0,\L/2,\L) -- (-0.14,\L/2+0.1,\L+0.14);
    \foreach \t in {0.12, 0.38, 0.62, 0.88} {
        \node[sample coarse=purple!70] at ({\L*(1-\t)}, {\t*\L/2}, {\t*\L}) {};
    }
    \foreach \t in {0.35, 0.48} {
        \node[sample refined=purple!70] at ({\L*(1-\t)}, {\t*\L/2}, {\t*\L}) {};
    }
\end{scope}


\begin{scope}[shift={(\colD + 0.8, \rowTop)},
    x={(-0.65cm,-0.95cm)}, y={(1.6cm,0cm)}]
    
    \def\S{2.6}
    \draw[box2d] (0,0) -- (\S,0) -- (\S,\S) -- (0,\S) -- cycle;
    
    \def\lx{1.3}
    \def\ly{1.3}
    \draw[red!80!black, thick] (\lx,\ly) ellipse (0.14 and 0.12);
    \draw[red!60!black, thick] (\lx,\ly) ellipse (0.30 and 0.25);
    \draw[red!40!black, thick] (\lx,\ly) ellipse (0.48 and 0.40);
    \draw[red!25!black, thick] (\lx,\ly) ellipse (0.70 and 0.58);
    \node[red!60!black, font=\small] at (\lx+0.18, \ly+0.78) {$\mathcal{L}$};
    
    \def\cx{1.3}
    \def\cy{1.3}
    \draw[ray=orange!75!black] (\cx,\cy) -- (\S+0.18, \S+0.12);
    \foreach \t in {0.4, 0.75} {
        \node[sample refined=orange!75!black] at ({\cx + \t*(\S-\cx)}, {\cy + \t*(\S-\cy)}) {};
    }
    \draw[ray=orange!75!black] (\cx,\cy) -- (\S+0.18, 0.7);
    \foreach \t in {0.45, 0.8} {
        \node[sample refined=orange!75!black] at ({\cx + \t*(\S-\cx)}, {\cy + \t*(0.7-\cy)}) {};
    }
    \draw[ray=orange!75!black] (\cx,\cy) -- (0.12, -0.18);
    \foreach \t in {0.5, 0.85} {
        \node[sample coarse=orange!75!black] at ({\cx + \t*(0.18-\cx)}, {\cy + \t*(0-\cy)}) {};
    }
    \fill[orange!75!black] (\cx,\cy) circle (2.5pt);
\end{scope}

\node[font=\bfseries\Large] at (\colD + 2.0, \rowBot - 1.8) {Sunburst};

\begin{scope}[shift={(\colD + 0.8, \rowBot)},
    x={(-0.65cm,-0.48cm)}, y={(1.65cm,0cm)}, z={(0cm,1.65cm)}]
    
    \draw[box3d back] (0,0,0) -- (\L,0,0);
    \draw[box3d back] (0,0,0) -- (0,\L,0);
    \draw[box3d back] (0,0,0) -- (0,0,\L);
    \draw[box3d back] (\L,0,0) -- (\L,\L,0);
    \draw[box3d back] (0,\L,0) -- (\L,\L,0);
    \draw[box3d back] (\L,0,0) -- (\L,0,\L);
    \draw[box3d back] (0,\L,0) -- (0,\L,\L);
    \draw[box3d back] (0,0,\L) -- (\L,0,\L);
    \draw[box3d back] (0,0,\L) -- (0,\L,\L);
    
    \shade[ball color=red!60!black, opacity=0.6] (1.3,1.3,1.3) circle (0.38);
    
    \draw[box3d] (\L,\L,0) -- (\L,\L,\L);
    \draw[box3d] (\L,0,\L) -- (\L,\L,\L);
    \draw[box3d] (0,\L,\L) -- (\L,\L,\L);
    
    \def\cx{1.3}
    \def\cy{1.3}
    \def\cz{1.3}
    \draw[ray=orange!75!black] (\cx,\cy,\cz) -- (\L+0.14,\L+0.1,\L+0.1);
    \foreach \t in {0.4, 0.75} {
        \node[sample refined=orange!75!black] at ({\cx + \t*(\L-\cx)}, {\cy + \t*(\L-\cy)}, {\cz + \t*(\L-\cz)}) {};
    }
    \draw[ray=orange!75!black] (\cx,\cy,\cz) -- (\L+0.14,0.65,0.5);
    \foreach \t in {0.45, 0.8} {
        \node[sample refined=orange!75!black] at ({\cx + \t*(\L-\cx)}, {\cy + \t*(0.7-\cy)}, {\cz + \t*(0.55-\cz)}) {};
    }
    \draw[ray=orange!75!black] (\cx,\cy,\cz) -- (0.08,-0.14,0.15);
    \foreach \t in {0.5, 0.85} {
        \node[sample coarse=orange!75!black] at ({\cx + \t*(0.18-\cx)}, {\cy + \t*(0-\cy)}, {\cz + \t*(0.25-\cz)}) {};
    }
    \fill[orange!75!black] (\cx,\cy,\cz) circle (2.5pt);
\end{scope}

\begin{scope}[shift={(2, \rowBot - 6.5)}]
    \draw[red!80!black, very thick] (0,0) ellipse (0.22 and 0.17);
    \draw[red!50!black, very thick] (0,0) ellipse (0.42 and 0.32);
    \node[font=\Large, anchor=west] at (0.7, 0) {Iso-likelihood $\mathcal{L}(\theta)$};
    
    \node[sample coarse=black!70, inner sep=2.5pt, line width=0.7pt] at (6.5, 0) {};
    \node[font=\Large, anchor=west] at (6.9, 0) {Coarse sample};
    
    \node[sample refined=black!70, inner sep=2.8pt] at (12.0, 0) {};
    \node[font=\Large, anchor=west] at (12.4, 0) {Refined sample};
    
    \draw[ray outside=black!70, line width=1.2pt] (17.2, 0) -- (18.2, 0);
    \node[font=\Large, anchor=west] at (18.5, 0) {Outside prior};
\end{scope}

\end{tikzpicture}%
}
\caption{Ray casting strategies for discovering likelihood mass. Top row: 2D projection showing iso-likelihood contours. Bottom row: 3D prior hypercube with likelihood region (red ellipsoid). \textbf{V2V}: vertex-to-vertex rays traverse full diagonals. \textbf{W2W}: wall-to-wall rays connect opposing faces. \textbf{V2E}: vertex-to-edge rays provide intermediate angles. \textbf{Sunburst}: rays radiate from the center in all directions. Open circles show coarse uniform sampling; filled circles show refined sampling concentrated near the likelihood.}
\label{fig:raycasting}
\end{figure}

\subsubsection{Two-Pass Sampling}

Along each ray $r(t) = p_{\text{start}} + t \cdot (p_{\text{end}} - p_{\text{start}})$ for $t \in [0,1]$:

\paragraph{Coarse Pass.} Sample uniformly at $n_{\text{coarse}}$ points. Evaluate log-likelihoods. Identify the ``active region'' where $\ell(\theta) > \ell_{\max} - \Delta$ for threshold $\Delta$.

\paragraph{Refinement Pass.} Resample with density proportional to likelihood within the active region. This concentrates evaluations where they matter.

\subsubsection{Scale Discovery via SingleWhip (\begin{CJK}{UTF8}{bsmi}單鞭\end{CJK})}

Before exploration, we estimate characteristic scales from ray samples using curvature analysis. For log-likelihood samples $\ell(t)$ along rays, we compute the second derivative at multiple skip-lengths $h$:
\begin{equation}
    \kappa(h) = \max_{t} \left| \frac{\ell(t+h) - 2\ell(t) + \ell(t-h)}{h^2} \right|
\end{equation}
Transitions where $\kappa(h)$ drops sharply indicate characteristic scales $\lambda_\text{fine}$, $\lambda_\text{mid}$, $\lambda_\text{coarse}$.

\subsubsection{ChiSao (\begin{CJK}{UTF8}{bsmi}黐手\end{CJK} -- Sticky Hands) Exploration}
\label{sec:ChiSao}

ChiSao implements the oscillating convergence-anticonvergence strategy. The name refers to the Wing Chun ``sticky hands'' technique---samples that find true peaks ``stick'' and remain frozen while others continue exploring.

\paragraph{The Oscillation Cycle.} For \texttt{n\_oscillations} cycles (default 3):

\begin{enumerate}
    \item \textbf{Convergence Phase}: Release all stuck masks and run batched L-BFGS toward local maxima for $n_{\text{converge}}$ iterations.
    
    \item \textbf{Stick Detection}: Samples satisfying \emph{both} conditions are marked ``stuck'':
    \begin{itemize}
        \item Gradient $L_\infty$ norm below tolerance (converged)
        \item Log-likelihood $\ell \geq \ell_{\max} + \log(0.1)$, i.e., likelihood at least 10\% of the current maximum
    \end{itemize}
    The likelihood threshold prevents sticking at low-quality local maxima.
    
    \item \textbf{Deduplication}: Peaks are deduplicated using the $L_\infty$ metric, removing $K_{\text{lost}}$ duplicate samples. The sample with highest likelihood survives; its inverse Hessian estimate is preserved for width estimation.
    
    \item \textbf{Repulse Monkey (\begin{CJK}{UTF8}{bsmi}倒攆猴\end{CJK}) / Golden Rooster (\begin{CJK}{UTF8}{bsmi}金雞獨立\end{CJK})}: If $K_{\text{lost}} > 0$ and not the last oscillation, maintain sample count:
    \begin{itemize}
        \item \textbf{Standard case} ($\geq 5$ unconverged samples): Repulse Monkey reseeds by shooting rays from unconverged samples in random directions
        \item \textbf{Near-exhaustion} ($< 5$ unconverged): Golden Rooster refills from \emph{converged} peaks using orthonormal ray directions (via QR decomposition), exploiting the GPU's parallel capacity
    \end{itemize}
    New samples are marked unstuck to enable exploration. Golden Rooster adaptively disables itself if the new samples don't discover any new peaks.
    
    \item \textbf{Hands Like Clouds (\begin{CJK}{UTF8}{bsmi}雲手\end{CJK})}: If not the last oscillation, unstuck samples take $n_{\text{cloud}}$ gradient ascent steps using a \emph{stochastically smoothed} gradient:
    \begin{equation}
        \nabla \ell_\sigma(\theta) \approx \frac{1}{K} \sum_{k=1}^{K} \nabla \ell(\theta + \sigma \cdot z_k), \quad z_k \sim \mathcal{N}(0, I_d)
    \end{equation}
    This blurs features smaller than $\sigma$, revealing global structure through local noise. For multi-scale problems (Rastrigin, Ackley), this allows samples to see the global basin through local ripples. The smoothing scale $\sigma$ is auto-estimated from sample spread if not provided.
    
    \item \textbf{Anti-convergence Phase}: Unstuck samples receive $n_{\text{anticonverge}}$ momentum-based gradient \emph{descent} steps:
    \begin{equation}
        v_{t+1} = \mu \cdot v_t - \alpha \cdot \nabla\ell(\theta_t), \quad \theta_{t+1} = \theta_t + v_{t+1}
    \end{equation}
    with momentum $\mu = 0.9$ and step size $\alpha$. The momentum allows samples to ``overshoot'' valleys and discover new peaks. Stuck samples remain frozen at their peaks throughout this phase.
\end{enumerate}

\paragraph{Design Rationale.} The ordering is deliberate:
\begin{itemize}
    \item \textbf{Deduplication after convergence}: Samples are most clustered after convergence, making duplicate detection most effective
    \item \textbf{Reseeding before smoothing}: New samples benefit from the Hands Like Clouds phase
    \item \textbf{Smoothing before anti-convergence}: Nudges samples toward global basins before the momentum-based exploration begins
    \item \textbf{Steps 4--6 skipped on last oscillation}: Final convergence produces the output peaks without further perturbation
\end{itemize}

\subsubsection{Algorithm Summary}

\begin{algorithm}[H]
\caption{CarryTiger: Mode Discovery via ChiSao Exploration}
\begin{algorithmic}[1]
\Require Ray samples $\{x_i\}$, log-likelihoods $\{\ell_i\}$, parameters $n_{\text{osc}}, n_{\text{conv}}, n_{\text{anti}}$
\Ensure Coarse peaks, RayBank

\State Initialize samples $x$ from top ray samples
\State $\texttt{stuck} \gets \texttt{False}^N$
\For{$k = 1, \ldots, n_{\text{osc}}$}
    \State $\texttt{stuck} \gets \texttt{False}^N$ \Comment{Release all}
    \State $x, \nabla\ell \gets \texttt{L-BFGS}(x, n_{\text{conv}})$ \Comment{Convergence}
    \State $\texttt{stuck} \gets (\|\nabla\ell\|_\infty < \epsilon) \land (\ell \geq \ell_{\max} + \log 0.1)$ \Comment{Stick detection}
    \State $x \gets \texttt{deduplicate}(x, L_\infty)$ \Comment{Remove duplicates}
    \If{$k < n_{\text{osc}}$}
        \State $x \gets \texttt{reseed}(x, \texttt{stuck})$ \Comment{Repulse Monkey / Golden Rooster}
        \State $x[\lnot\texttt{stuck}] \gets \texttt{HLC}(x[\lnot\texttt{stuck}])$ \Comment{Hands Like Clouds}
        \State $x[\lnot\texttt{stuck}] \gets \texttt{anti-converge}(x[\lnot\texttt{stuck}], n_{\text{anti}})$ \Comment{Momentum descent}
    \EndIf
\EndFor
\State \Return $x[\texttt{stuck}]$, RayBank
\end{algorithmic}
\end{algorithm}

\subsubsection{Output}

Module 1 returns $K$ coarse peak locations with width estimates, plus RayBank containing all ray samples and log-likelihoods.

\subsection{Module 2: GreenDragon (\begin{CJK}{UTF8}{bsmi}青龍出水\end{CJK} -- Green Dragon Rises from Water)}
\label{sec:GreenDragon}

GreenDragon refines coarse peak locations to machine precision.

\subsubsection{Seeding Strategy}

For each coarse peak at $\theta^*$ with estimated width $w$:
\begin{itemize}
    \item \textbf{Default mode}: Seed $2d$ samples at $\theta^* \pm 0.9w \cdot e_i$ for each axis $i$
    \item \textbf{Fast mode}: Seed 20 random perturbations
\end{itemize}

The axis-aligned seeds enable GPU-parallel diagonal Hessian computation. All $2d$ perturbations are evaluated in a single batched GPU call:
\begin{equation}
    H_{ii} = \frac{\ell(\theta^* + h e_i) - 2\ell(\theta^*) + \ell(\theta^* - h e_i)}{h^2}
\end{equation}
This yields the full diagonal Hessian at $O(1)$ wall-clock cost (given $P \gg d$ GPU cores), compared to $O(d)$ for sequential evaluation.

\subsubsection{Batched L-BFGS Refinement}

All seeds across all peaks optimize simultaneously. Iteration count adapts: $\max(10, 3\log_2(d))$.

\subsubsection{Trajectory Storage}

L-BFGS trajectories are stored in the TrajectoryBank. Each trajectory records the sequence of positions visited during optimization, enabling Module 3 (BendTheBow) to analyze optimization paths for rotation detection without additional likelihood evaluations. This is the key mechanism for zero-cost geometry detection.

\subsubsection{Filtering: Saddle Points and Non-Stationary Points}

Two filters ensure only true maxima survive:

\paragraph{Saddle Point Filter.} Candidates with any $H_{ii} > 0$ (positive curvature indicates an ascending direction) are saddle points and removed.

\paragraph{Gradient Filter.} Candidates with $\|\nabla\ell\|_\infty > \epsilon_{\text{grad}}$ are not truly stationary and removed. The gradient norm is available ``for free'' from L-BFGS---no additional likelihood evaluations required.

\subsubsection{Algorithm Summary}

\begin{algorithm}[H]
\caption{GreenDragon: Peak Refinement}
\begin{algorithmic}[1]
\Require Coarse peaks $\{\theta^*_k\}$ with widths $\{w_k\}$ from Module 1
\Ensure Refined peaks, diagonal Hessians, TrajectoryBank

\For{each peak $k$} \Comment{GPU-parallel across all peaks}
    \State Seed $2d$ samples at $\theta^*_k \pm 0.9 w_k \cdot e_i$ for $i = 1, \ldots, d$
\EndFor
\State $x, \nabla\ell, \texttt{trajectories} \gets \texttt{L-BFGS}(\text{all seeds}, \text{return\_trajectories}=\texttt{True})$
\State Store trajectories in TrajectoryBank
\State Compute diagonal Hessian $H_{ii}$ from seed evaluations \Comment{Already computed}
\State $\texttt{is\_max} \gets \texttt{all}(H_{ii} < 0, \text{axis}=1)$ \Comment{Saddle filter}
\State $\texttt{is\_stationary} \gets \|\nabla\ell\|_\infty < \epsilon_{\text{grad}}$ \Comment{Gradient filter (free)}
\State $\texttt{keep} \gets \texttt{is\_max} \land \texttt{is\_stationary}$
\State Deduplicate surviving peaks
\State \Return refined peaks, $\{H_{ii}\}$, TrajectoryBank
\end{algorithmic}
\end{algorithm}

\subsubsection{Output}

Module 2 returns $K'$ refined peaks with diagonal Hessians $\{H_{ii}\}$, plus the TrajectoryBank containing all L-BFGS optimization paths. The TrajectoryBank is the primary input for rotation detection in Module 3 (BendTheBow).

\subsection{Module 3: BendTheBow (\begin{CJK}{UTF8}{bsmi}彎弓射虎\end{CJK} -- Bend the Bow, Shoot the Tiger)}
\label{sec:BendTheBow}

BendTheBow computes the Bayesian evidence via Laplace approximation.

\subsubsection{Laplace Approximation}

For each refined peak $\theta^*_k$ with Hessian $H_k$:
\begin{equation}
    \log Z_k = \ell(\theta^*_k) + \frac{d}{2}\log(2\pi) - \frac{1}{2}\log|\det(-H_k)|
\end{equation}

This is exact for Gaussian posteriors and provides excellent approximations for near-Gaussian posteriors.

\subsubsection{Automatic Geometry Detection}

A key challenge is detecting when the posterior has off-diagonal correlations requiring a full Hessian. Module 2 provides only the diagonal $\{H_{ii}\}$ at $O(d)$ cost; the full Hessian costs $O(d^2)$.

\sunburst{} uses two-step detection:

\paragraph{Step 1: Perpendicular Step Fraction (Zero Extra Evaluations).}
Analyze L-BFGS trajectories from GreenDragon's TrajectoryBank. In whitened coordinates (scaled by $\sqrt{-H_{ii}}$), compute for each trajectory step:
\begin{equation}
    f_\perp = \frac{\|\Delta\theta_\perp\|}{\|\Delta\theta\|}
\end{equation}
where $\Delta\theta_\perp$ is the component perpendicular to the gradient direction. For axis-aligned posteriors, optimization proceeds along coordinate axes ($f_\perp \approx 0$). Significant perpendicular motion ($f_\perp > \epsilon_{\text{rot}}$) indicates rotation.

\paragraph{Step 2: Finite-Difference Probe (2 Evaluations).}
If trajectory-based detection is inconclusive, probe the off-diagonal structure directly:
\begin{equation}
    b = \frac{v^T H v - a}{d-1}, \quad v = \frac{1}{\sqrt{d}}[1,1,\ldots,1]^T
\end{equation}
where $a = \text{mean}(H_{ii})$. Significant $|b|$ indicates uniform correlation.

\paragraph{Full Hessian Computation.}
If rotation is detected, compute via central differences:
\begin{equation}
    H_{ij} = \frac{\ell_{++} - \ell_{+-} - \ell_{-+} + \ell_{--}}{4\epsilon^2}
\end{equation}
This requires $O(d^2)$ likelihood evaluations. On GPU with $P \gg d^2$ cores, all $2d^2$ perturbations are evaluated in a single batched call, yielding $O(d)$ wall-clock time.

\paragraph{Typical vs.\ Worst-Case Complexity.}
While the worst-case cost of computing a full Hessian scales as $O(d^2)$, \sunburst{} avoids this in the typical case by detecting axis-aligned structure and computing only diagonal curvature, triggering the full Hessian only when rotation is detected. This conditional approach yields:
\begin{equation}
    \text{Hessian cost} = \begin{cases} 
        O(d) & \text{axis-aligned posteriors (typical)} \\ 
        O(d^2) & \text{rotated posteriors (rare)}
    \end{cases}
\end{equation}

\subsubsection{Evidence Calculation}

\paragraph{Diagonal Hessian (Axis-Aligned Posteriors).}
\begin{equation}
    \log Z_k = \ell(\theta^*_k) + \frac{d}{2}\log(2\pi) - \frac{1}{2}\sum_{i=1}^{d} \log(-H_{ii})
\end{equation}

\paragraph{Full Hessian (General Posteriors).}
Compute via GPU-accelerated eigendecomposition:
\begin{equation}
    \log|\det(-H)| = \sum_{i=1}^{d} \log(\lambda_i)
\end{equation}
where $\{\lambda_i\}$ are eigenvalues of $-H$.

\subsubsection{Multimodal Evidence}

The total evidence sums over all $K$ modes:
\begin{equation}
    \log Z = \text{logsumexp}(\log Z_1, \ldots, \log Z_K)
\end{equation}

\subsubsection{Algorithm Summary}

\begin{algorithm}[H]
\caption{BendTheBow Evidence Calculation}
\begin{algorithmic}[1]
\Require Peaks $\{\theta^*_k\}$, diagonal Hessians $\{H_k\}$ from GreenDragon, TrajectoryBank from GreenDragon
\Ensure Log-evidence $\log Z$

\For{each peak $k = 1, \ldots, K$}
    \State $\texttt{rotated} \gets \texttt{perpendicular\_step\_fraction}(\text{trajectories}_k)$
    \If{inconclusive}
        \State $\texttt{rotated} \gets \texttt{probe\_offdiag}(\theta^*_k, H_k)$ \Comment{2 evals}
    \EndIf
    \If{rotated}
        \State $H_k \gets \texttt{full\_hessian}(\theta^*_k)$ \Comment{$O(d^2)$ evals}
        \State $\log Z_k \gets \ell_k + \frac{d}{2}\log(2\pi) - \frac{1}{2}\log|\det(-H_k)|$
    \Else
        \State $\log Z_k \gets \ell_k + \frac{d}{2}\log(2\pi) - \frac{1}{2}\sum_i \log(-H_{ii})$
    \EndIf
\EndFor
\State \Return $\texttt{logsumexp}(\log Z_1, \ldots, \log Z_K)$
\end{algorithmic}
\end{algorithm}

\subsection{Module 4: GraspBirdsTail (\begin{CJK}{UTF8}{bsmi}攬雀尾\end{CJK} -- Grasp Bird's Tail) [Optional]}
\label{sec:GraspBirdsTail}

GraspBirdsTail prepares a compacted likelihood for handoff to downstream samplers (PolyChord, dynesty, emcee). This module is optional and runs after evidence calculation when full posterior sampling is desired.

\subsubsection{Eigendecomposition Analysis}

Using the full Hessian $H$ at the dominant peak (or an approximation constructed from the trajectory bank), compute the eigendecomposition:
\begin{equation}
    -H = V \Lambda V^T, \quad \Lambda = \text{diag}(\lambda_1, \ldots, \lambda_d)
\end{equation}

The eigenvalues reveal the posterior geometry and classify each direction:
\begin{itemize}
    \item \textbf{Informative directions}: $\lambda_i > \epsilon_{\text{inform}}$ --- strongly constrained by data
    \item \textbf{Nuisance directions}: $\epsilon_{\text{nuisance}} < \lambda_i < \epsilon_{\text{inform}}$ --- weakly constrained
    \item \textbf{Degenerate directions}: $\lambda_i < \epsilon_{\text{nuisance}}$ --- unconstrained (flat)
\end{itemize}

\subsubsection{Dimensional Reduction}

The effective dimension $d_{\text{eff}}$ counts informative directions:
\begin{equation}
    d_{\text{eff}} = \#\{i : \lambda_i > \epsilon_{\text{inform}}\}
\end{equation}

The rotation matrix $V$ transforms to principal coordinates where:
\begin{itemize}
    \item Informative directions become the first $d_{\text{eff}}$ coordinates
    \item Nuisance directions can be marginalized analytically (Gaussian integral)
    \item Degenerate directions contribute their prior width
\end{itemize}

\subsubsection{Compacted Likelihood}

The reduced likelihood $\ell_{\text{red}}(\phi)$ operates on only $d_{\text{eff}}$ dimensions:
\begin{equation}
    \ell_{\text{red}}(\phi) = \ell(V_{\text{info}}^T \phi + \theta^*_{\text{nuisance}})
\end{equation}
where $V_{\text{info}}$ contains the informative eigenvectors and $\theta^*_{\text{nuisance}}$ fixes nuisance parameters at their MAP values.

\subsubsection{Algorithm Summary}

\begin{algorithm}[H]
\caption{GraspBirdsTail: Dimensional Reduction for Sampler Handoff}
\begin{algorithmic}[1]
\Require Dominant peak $\theta^*$, Hessian $H$, thresholds $\epsilon_{\text{inform}}, \epsilon_{\text{nuisance}}$
\Ensure Reduced likelihood, rotation matrix, marginal contributions

\State $V, \Lambda \gets \texttt{eig}(-H)$ \Comment{Eigendecomposition}
\State $\texttt{inform} \gets \{i : \lambda_i > \epsilon_{\text{inform}}\}$
\State $\texttt{nuisance} \gets \{i : \epsilon_{\text{nuisance}} < \lambda_i \leq \epsilon_{\text{inform}}\}$
\State $\texttt{degen} \gets \{i : \lambda_i \leq \epsilon_{\text{nuisance}}\}$
\State $d_{\text{eff}} \gets |\texttt{inform}|$
\State $V_{\text{info}} \gets V[:, \texttt{inform}]$
\State $\ell_{\text{red}}(\phi) \gets \ell(V_{\text{info}}^T \phi + \theta^*_{\text{nuisance}})$
\State $\log Z_{\text{nuisance}} \gets \sum_{i \in \texttt{nuisance}} \frac{1}{2}\log(2\pi/\lambda_i)$
\State $\log Z_{\text{degen}} \gets \sum_{i \in \texttt{degen}} \log(\Delta_i)$
\State \Return $\ell_{\text{red}}$, $V$, $d_{\text{eff}}$, $\log Z_{\text{nuisance}} + \log Z_{\text{degen}}$
\end{algorithmic}
\end{algorithm}

\subsubsection{Output}

GraspBirdsTail exports:
\begin{itemize}
    \item Reduced likelihood $\ell_{\text{red}}(\phi)$ on $d_{\text{eff}}$ dimensions
    \item Rotation matrix $V$ for coordinate transformation
    \item Bounds in the reduced space
    \item Marginal contributions from nuisance/degenerate directions
\end{itemize}

This enables efficient posterior sampling even when the original problem has many nuisance directions, as downstream samplers operate in the reduced $d_{\text{eff}}$-dimensional space.

\subsection{Scaling Strategies}
\label{sec:scaling}

The curse of dimensionality cannot be defeated---it is a fundamental property of high-dimensional spaces. However, \sunburst{} employs several strategies that delay its onset and mitigate its effects:

\subsubsection{Radial Integration from Peaks}

We integrate around detected modes, not over the full prior. The exponentially large prior volume enters only as an analytic normalizing constant.

\subsubsection{GPU Parallelization}

It is important to distinguish \emph{computational complexity} (number of likelihood evaluations) from \emph{wall-clock scaling} (time on parallel hardware):

\begin{center}
\begin{tabular}{l|c|c|c}
\textbf{Operation} & \textbf{Evaluations} & \textbf{CPU Wall-clock} & \textbf{GPU Wall-clock} \\
\hline
Gradient (finite diff.) & $2d$ & $O(d)$ & $O(1)$ \\
Diagonal Hessian & $2d$ & $O(d)$ & $O(1)$ \\
Full Hessian & $2d^2$ & $O(d^2)$ & $O(d)$ \\
L-BFGS iteration & $O(d)$ & $O(d)$ & $O(1)$ \\
\end{tabular}
\end{center}

With sufficient GPU parallelism, the $2d^2$ perturbations for a full Hessian can be evaluated in a small number of large batched calls. The \emph{algorithmic} complexity remains $O(d^2)$, but the \emph{wall-clock} time can scale closer to $O(d)$ over the regime where GPU throughput is saturated, and kernel-launch overhead is amortized.

\subsubsection{$O(K)$ Mode Scaling}

Nested sampling's theoretical scaling is $O(K \cdot d)$ live points, but empirical performance degrades far more severely with dimension due to the difficulty of constrained prior sampling~\cite{ashton2022nested,handley2015polychord}. \sunburst{} treats modes independently via local Laplace approximation, requiring only $O(K)$ effective samples regardless of dimension.

\subsubsection{Sample Reuse}

Every likelihood evaluation is banked. Module 3 reuses GreenDragon's TrajectoryBank for geometry detection, achieving zero-cost rotation diagnosis in the common case.

\subsection{Computational Complexity}

\begin{center}
\begin{tabular}{l|c|c}
\textbf{Method} & \textbf{Hessian Cost} & \textbf{Notes} \\
\hline
Laplace (diagonal) & $O(d)$ & Axis-aligned only \\
Laplace (full) & $O(d^2)$ & General posteriors \\
Riemannian HMC & $O(d^2)$ per step & Metric update every step \\
Nested Sampling & $O(d^2)$ implicit & Live point proposal \\
\textbf{\sunburst{} (typical)} & $O(d)$ & Rotation detection avoids full Hessian \\
\textbf{\sunburst{} (worst-case)} & $O(d^2)$ & Rotated posteriors \\
\end{tabular}
\end{center}

For multimodal problems with $K$ modes:

\begin{center}
\begin{tabular}{l|c|c}
\textbf{Method} & \textbf{Evaluations} & \textbf{GPU Wall-clock} \\
\hline
Nested Sampling & $O(K \cdot d^2 \cdot H)$ & $O(K \cdot d^2 \cdot H)$ \\
\textbf{\sunburst{} (typical)} & $O(K \cdot d)$ & $O(K \cdot d^{0.5})$ \\
\textbf{\sunburst{} (rotated)} & $O(K \cdot d^2)$ & $O(K \cdot d)$ \\
\end{tabular}
\end{center}

where $H$ is the information gain (typically $\propto d$). The empirical $O(d^{0.5})$ wall-clock scaling reflects GPU parallelization of the $O(d)$ algorithmic complexity.

Note that nested sampling's frequently cited scaling expressions can understate end-to-end cost in high dimension. In practical implementations, each iteration requires \emph{constrained prior sampling} within an evolving likelihood contour, together with geometric updates (e.g.\ covariance estimation and bound construction) that are at least quadratic in $d$. As dimensionality increases, the difficulty of proposing valid constrained samples typically dominates and can induce substantially worse-than-quadratic scaling in wall-clock time. Skilling~\cite{skilling2006nested} gives an idealized $O(d^2)$ cost term, whereas Handley et al.~\cite{handley2015polychord} report $O(d^3)$ scaling for PolyChord in realistic regimes; the review by Ashton et al.~\cite{ashton2022nested} discusses the persistent gap between asymptotic descriptions and practical performance across implementations.

\paragraph{Accuracy Concerns.}
Evidence estimates from nested sampling are sensitive to the number of live points, proposal/bounding strategy, and termination criterion, and their reported internal uncertainties need not be well calibrated under finite budgets. Even on Gaussian test problems, published results report non-negligible evidence uncertainty (e.g.\ $\sim\!0.4$ in $\log Z$ units under typical PolyChord settings)~\cite{handley2015polychord}. Buchner~\cite{buchner2014statistical} developed diagnostic tests indicating systematic bias in nested-sampling evidence estimation; Higson et al.~\cite{higson2018sampling} document implementation-specific failure modes in which posterior exploration is incomplete, producing biased evidence estimates; and Nelson et al.~\cite{nelson2020quantifying} report dimension-dependent biases in MultiNest. The JAXNS benchmarks~\cite{albert2023jaxns} further show MultiNest failing at moderate dimension on Gaussian likelihoods under standard configurations.

In contrast, \sunburst{} targets calibrated evidence estimation directly for Gaussian and near-Gaussian posteriors. On exactly Gaussian posteriors (including off-center, cigar, correlated, and rotated-cigar geometries), the Laplace approximation is exact and \sunburst{} attains numerical agreement at floating-point tolerance through 512D--1024D in our benchmarks. On Gaussian mixtures, accuracy depends on configuration: with conservative settings (3 oscillations and full GreenDragon refinement) the 4-mode mixture yields $0.07\%$ error at 512D, while the fast configuration trades accuracy for speed (e.g.\ $1.4\%$ error at 768D).

More broadly, posterior sampling and evidence calibration impose different algorithmic requirements; accuracy in posterior exploration does not automatically imply calibrated $\log Z$ at high dimension under practical compute budgets.

\subsection{Limitations}

The Laplace approximation assumes the posterior is well-approximated by a Gaussian around each mode. This is exact for Gaussian posteriors and accurate for many physics applications. For posteriors with heavy tails, strong skewness, or banana-shaped degeneracies, the Laplace approximation may introduce bias. 

Current limitations include:
\begin{itemize}
    \item \textbf{Prior containment assumed}: The likelihood must be well-contained within the prior hypercube; truncation by prior boundaries is not currently handled
    \item \textbf{Uniform priors only}: Non-uniform or informative priors require future extensions
    \item \textbf{Non-Gaussian posteriors}: Heavy tails and skewness not captured by Laplace
    \item \textbf{Banana-shaped degeneracies}: Curved ridges violate local Gaussian assumption
    \item \textbf{Phase transitions}: Sharp features in the likelihood may be missed by ray casting
\end{itemize}

Extending \sunburst{} to handle such cases---via importance sampling corrections, prior-aware integration, or hybrid methods---remains an area for future work (Section~\ref{sec:future}).


\section{Theoretical Foundations}
\label{sec:theory}

This section establishes the theoretical foundations for \sunburst{}'s three key innovations: radial ray mode discovery, Laplace approximation for evidence calculation, and GPU parallelization complexity.

\subsection{Mode Discovery via Radial Rays}
\label{sec:theory-rays}

The first stage of \sunburst{} discovers posterior modes by casting rays from the prior boundary toward the interior. We formalize when this approach is guaranteed to locate modes.

\paragraph{Log-concavity.}
A probability density $p(\theta)$ is \emph{log-concave} if $\log p(\theta)$ is a concave function, i.e., for all $\theta_1, \theta_2$ and $\lambda \in [0,1]$:
\begin{equation}
    \log p(\lambda\theta_1 + (1-\lambda)\theta_2) \geq \lambda \log p(\theta_1) + (1-\lambda) \log p(\theta_2)
\end{equation}
Many common posterior distributions are log-concave, including Gaussian, exponential, logistic, and Laplace distributions, as well as posteriors arising from generalized linear models with log-concave priors \citep{prekopa1973logarithmic}.

\paragraph{Radial ray mode detection.}
Let $p(\theta)$ be a unimodal log-concave density on a convex domain $\Omega \subset \mathbb{R}^d$ with unique mode at $\theta^*$. For any boundary point $p \in \partial\Omega$ and unit direction $\hat{d}$ pointing into $\Omega$, the ray $r(t) = p + t\hat{d}$ satisfies: (i) the log-likelihood profile $\ell(t) = \log p(r(t))$ is unimodal in $t$, and (ii) the maximum occurs at the value $t^*$ maximizing $\ell(t)$, where $r(t^*)$ lies on the line segment connecting $p$ to $\theta^*$.

This follows because the restriction of a concave function to a line is concave. Since $\log p$ is concave on $\Omega$, the function $\ell(t) = \log p(r(t))$ is concave in $t$ over the interval where $r(t) \in \Omega$. A concave function on a convex domain has at most one local maximum, which must be the global maximum.

For multimodal posteriors, the log-concavity assumption fails globally but typically holds locally around each mode. This motivates the local maximum detection in the CarryTiger algorithm (Section~\ref{sec:CarryTiger}): each ray may exhibit multiple local maxima corresponding to distinct modes traversed, which are identified by finding points where the likelihood gradient along the ray changes sign.

These guarantees do not extend to posteriors with disconnected support, narrow ridges aligned orthogonally to ray directions, or highly oscillatory likelihood structure; ChiSao’s stochastic reseeding addresses such cases empirically but without formal completeness guarantees.

\paragraph{Non-convex posteriors.}
When the posterior is not globally log-concave, radial rays may exhibit multiple peaks. \sunburst{} handles this through ChiSao oscillation (Section~\ref{sec:ChiSao}), which randomizes ray directions across multiple passes to ensure coverage of all significant modes.

\subsection{Laplace Approximation for Evidence}
\label{sec:theory-laplace}

Given a mode location $\theta^*$, \sunburst{} computes the evidence contribution using the Laplace approximation. 

\paragraph{The Laplace approximation.}
Let $p(\theta|D) \propto \exp(-nf(\theta))$ where $f$ is smooth and has a unique minimum at $\theta^*$ with positive-definite Hessian $H = \nabla^2 f(\theta^*)$. Then \citep{tierney1986accurate}:
\begin{equation}
    \int_{\mathbb{R}^d} \exp(-nf(\theta))\, d\theta = \exp(-nf(\theta^*)) \cdot \frac{(2\pi)^{d/2}}{n^{d/2} |H|^{1/2}} \cdot \left(1 + O(n^{-1})\right)
\end{equation}

Taking logarithms yields the evidence approximation used in \sunburst{}:
\begin{equation}
    \log \mathcal{Z} \approx \log \mathcal{L}(\theta^*) + \frac{d}{2}\log(2\pi) - \frac{1}{2}\log|\mathbf{H}|
    \label{eq:laplace-evidence}
\end{equation}
where $\mathbf{H} = -\nabla^2 \log \mathcal{L}(\theta^*)$ is the observed information matrix (negative Hessian of log-likelihood at the mode).

\paragraph{Exactness for Gaussian posteriors.}
When $\log \mathcal{L}(\theta) = -\frac{1}{2}(\theta - \mu)^\top \Sigma^{-1} (\theta - \mu) + c$, the Laplace approximation is \emph{exact}: all higher-order terms in the Taylor expansion vanish identically. This explains \sunburst{}'s machine-precision accuracy on Gaussian test functions.

\paragraph{Error bounds.}
For non-Gaussian posteriors, the $O(n^{-1})$ error term depends on the third and fourth derivatives of $f$ at the mode \citep{kass1995bayes}. Posteriors with small skewness and kurtosis (near-Gaussian shape) yield smaller errors.

\paragraph{Multimodal extension.}
For a posterior with $K$ well-separated modes at $\{\theta^*_k\}_{k=1}^K$, the total evidence decomposes as:
\begin{equation}
    \mathcal{Z} = \sum_{k=1}^K \mathcal{Z}_k \approx \sum_{k=1}^K \exp\left(\log \mathcal{L}(\theta^*_k) + \frac{d}{2}\log(2\pi) - \frac{1}{2}\log|\mathbf{H}_k|\right)
    \label{eq:mixture-evidence}
\end{equation}
This mixture-of-Laplace approximation is accurate when modes are sufficiently separated that their Gaussian approximations have negligible overlap, a condition typically satisfied for well-defined multimodal posteriors.

\subsection{Computational Complexity Analysis}
\label{sec:theory-complexity}

We analyze \sunburst{}'s computational complexity and explain its observed sub-linear scaling.

\paragraph{Traditional nested sampling complexity.}
Nested sampling's computational cost is dominated by the \emph{constrained sampling problem}: at each iteration, a new point must be drawn uniformly from the prior subject to the likelihood constraint $\mathcal{L}(\theta) > \mathcal{L}_{\min}$. This constrained region shrinks by a factor $\approx e^{-1/N}$ at each iteration, where $N$ is the number of live points.

The total number of iterations required scales as $O(N \cdot |\log \mathcal{Z}|/\epsilon)$ to achieve relative evidence error $\epsilon$ \citep{skilling2006nested}. The critical question is how the cost of each constrained sample scales with dimension $d$.

\textbf{Ellipsoidal methods} (MultiNest): The algorithm fits bounding ellipsoids to the live point cloud, requiring $O(d^2)$ operations for covariance estimation. Proposals are drawn uniformly from these ellipsoids and accepted if they satisfy the likelihood constraint. The acceptance rate degrades as the ellipsoid approximation becomes poor in high dimensions, and the number of live points must scale as $N = O(d \cdot K)$ to capture multimodal structure \citep{feroz2009multinest}.

\textbf{Slice sampling methods} (PolyChord): Each constrained sample requires $O(d)$ slice expansions, with each expansion involving multiple likelihood evaluations to find the slice boundaries \citep{handley2015polychord}. Crucially, these $d$ slice operations are \emph{inherently sequential}---each slice depends on the result of the previous one---making GPU parallelization ineffective.

Combining these factors, the total complexity becomes:
\begin{equation}
    T_{\text{nested}} = O\left(\frac{N \cdot |\log \mathcal{Z}|}{\epsilon} \cdot C_{\text{sample}}(d)\right) = O\left(\frac{d \cdot K \cdot |\log \mathcal{Z}|}{\epsilon} \cdot d\right) = O\left(\frac{d^2 \cdot K \cdot |\log \mathcal{Z}|}{\epsilon}\right)
\end{equation}
where the $d^2$ factor arises from $N \propto d$ live points and $O(d)$ cost per constrained sample. This explains why PolyChord and dynesty become intractable beyond $d \approx 30$ in our benchmarks.

\paragraph{\sunburst{} complexity.}
\sunburst{}'s three stages have distinct complexity profiles:

\textbf{Mode discovery (CarryTiger):} Casts $N_{\text{rays}}$ rays with $N_{\text{steps}}$ evaluations each. Total: $O(N_{\text{rays}} \cdot N_{\text{steps}})$ likelihood calls, independent of dimension when parallelized.
    
\textbf{Peak refinement (GreenDragon):} L-BFGS optimization requiring $O(d)$ gradient evaluations per iteration, with $O(\log(1/\epsilon))$ iterations for convergence. Each gradient uses $2d$ finite-difference evaluations. Total per mode: $O(d \cdot \log(1/\epsilon))$.
    
\textbf{Evidence calculation (BendTheBow):} Hessian computation via finite differences requires $O(d^2)$ evaluations; eigendecomposition costs $O(d^3)$ floating-point operations. Total per mode: $O(d^2)$ likelihood calls.

\paragraph{GPU parallelization effect.}
Let $P$ denote the number of GPU parallel processors (typically $P > 10^4$). For $N$ independent likelihood evaluations:
\begin{equation}
    T_{\text{wall}} = \left\lceil \frac{N}{P} \right\rceil \cdot T_{\text{single}} \approx \frac{N}{P} \cdot T_{\text{single}}
\end{equation}
When $N < P$, all evaluations complete in a single parallel batch, yielding $O(1)$ wall-clock time regardless of $N$.

This transforms the complexity of each stage. Mode discovery: $O(N_{\text{rays}} \cdot N_{\text{steps}} / P) \approx O(1)$ for typical ray counts. Gradient computation: $2d$ evaluations parallelize to $O(\lceil 2d/P \rceil) = O(1)$ for $d < P/2$. Hessian computation: $O(d^2)$ evaluations yield $O(d^2/P)$ wall-clock time.

\paragraph{Effective scaling.}
For $K$ modes in dimension $d$ with $P$ processors:
\begin{equation}
    T_{\text{total}} = O\left(K \cdot \left(1 + \frac{d^2}{P}\right) + d^3/F\right)
\end{equation}
where $F$ is the GPU floating-point throughput for eigendecomposition. For $d \ll \sqrt{P}$ (satisfied for $d < 100$ on modern GPUs with $P > 10^4$), this reduces to O(K) in the pre-asymptotic GPU-saturated regime ($d^2 \ll P$), before Hessian cost dominates.

Our empirical measurements (Section~\ref{sec:results}) show $T \propto d^{0.52}$ over $4$--$1024$D, consistent with the Hessian computation ($d^2$ evaluations parallelized across $P$ processors) dominating once $d^2 > P$.

\paragraph{Mitigating dimensional scaling.}
Traditional methods scale as $O(K \cdot d^2)$ or worse due to the sequential nature of constrained sampling. \sunburst{}'s $O(K)$ scaling for moderate dimensions represents a significant complexity improvement enabled by the combination of (i) mode-centric rather than volume-based integration and (ii) massive GPU parallelism that converts sequential operations into parallel batches. The curse of dimensionality is not eliminated---the Hessian computation still costs $O(d^2)$---but its impact is substantially reduced compared to traditional nested sampling.

\subsection{Conditions for Accuracy}
\label{sec:theory-conditions}

\sunburst{}'s theoretical guarantees require several conditions:

\textbf{Mode discovery completeness:} All significant modes must be found. The ChiSao oscillation strategy addresses this through multi-directional ray casting, but pathologically hidden modes may be missed.
    
\textbf{Laplace approximation validity:} Each mode must be sufficiently Gaussian-like that the Laplace approximation provides adequate accuracy. Strongly non-Gaussian posteriors (heavy tails, curved degeneracies, saddle points) will incur larger errors.
    
\textbf{Mode separation:} The mixture-of-Laplace approximation~\eqref{eq:mixture-evidence} assumes negligible overlap between mode contributions. Highly overlapping modes require more sophisticated treatment.
    
\textbf{Bounded domain:} \sunburst{} operates on finite prior bounds. Improper priors or distributions with significant probability mass outside the bounds will yield biased evidence estimates.

When these conditions are satisfied---as they are for the Gaussian-like posteriors common in physical parameter estimation---\sunburst{} achieves relative error below $10^{-12}$ (double-precision tolerance)
 with sub-linear dimensional scaling. Section~\ref{sec:results} validates these theoretical predictions empirically.


\section{Results}
\label{sec:results}

\subsection{Accuracy Validation}

We evaluated \sunburst{} on isotropic Gaussian posteriors---the algorithm's native representation---across dimensions 2 to 1024. For these distributions, \sunburst{} achieved relative error below $10^{-12}$ (double precision limit) at all dimensions tested (Table~\ref{tab:scaling}).

This precision arises because the Laplace approximation is \emph{mathematically exact} for Gaussian distributions: the log-likelihood is exactly quadratic, so the second-order Taylor expansion introduces no error. The pipeline detects posterior geometry automatically---GreenDragon estimates diagonal Hessian structure during peak refinement, and BendTheBow analyzes optimization trajectories to detect rotation. For axis-aligned posteriors, the diagonal Hessian suffices; for rotated or correlated posteriors, the full Hessian is computed. In either case, per-peak whitening using the appropriate Hessian transforms the posterior to unit-Gaussian form, and the Laplace integral becomes exact.

\subsection{Wall-Clock Scaling}

\begin{figure}[htbp]
\centering
\includegraphics[width=\textwidth]{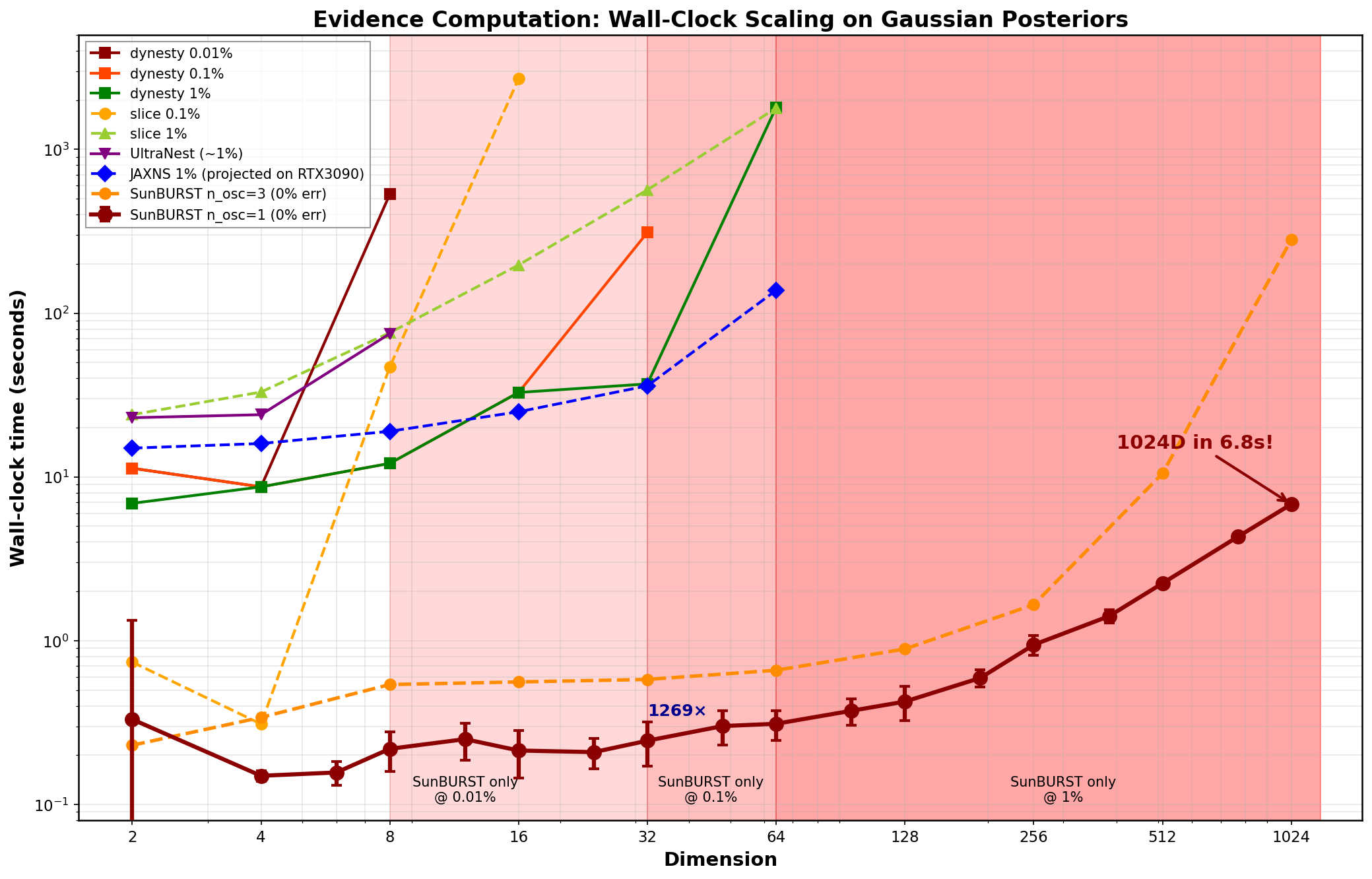}
\caption{Wall-clock scaling comparison on Gaussian posteriors. \sunburst{} (dark red circles) maintains near-constant runtime through 64 dimensions before transitioning to $O(d^{1.0})$ scaling. All competitors exhibit exponential growth: dynesty variants timeout beyond 32--64D, UltraNest beyond 8D. Shaded regions indicate dimensions where only \sunburst{} achieves each accuracy level. \sunburst{} achieves 1024D in 6.8 seconds.}
\label{fig:scaling}
\end{figure}

The central result of this work is a dramatic improvement in wall-clock performance that enables near-real-time evidence calculation for Gaussian-like posteriors. Figure~\ref{fig:scaling} and Table~\ref{tab:scaling} present runtime versus dimension, revealing two distinct scaling regimes.

In the \textbf{pre-asymptotic regime} ($d \leq 64$), wall-clock cost is effectively \emph{independent} of dimension. A 64-dimensional Gaussian posterior requires only 0.23 seconds---comparable to the 0.10 seconds needed for 2 dimensions. This remarkable flatness arises because GPU parallelization amortizes the per-dimension cost: when thousands of GPU threads process dimensions in parallel, adding more dimensions incurs negligible wall-clock overhead.

In the \textbf{post-asymptotic regime} ($d > 64$), dimensional scaling emerges with exponent $\alpha \approx 0.69$. Even in this regime, performance remains practical: 128D completes in 0.43 seconds, 256D in 0.94 seconds, 512D in 2.2 seconds, and \textbf{1024D in 6.8 seconds}---all with machine-precision accuracy.

The transition between regimes reflects working set size approaching GPU cache limits. Importantly, even the post-asymptotic scaling of $O(d^{1.0})$ is dramatically better than the $O(d^2)$ to $O(d^3)$ scaling exhibited by nested sampling competitors.

\subsection{Comparison with Existing Methods}

We benchmarked \sunburst{} against established nested sampling implementations: \texttt{dynesty} with ellipsoidal bounds, \texttt{dynesty-slice} using slice sampling (equivalent to \polychord{}'s algorithm), parallelized variants with 20 workers, \texttt{UltraNest} (reactive nested sampling), and \texttt{JAXNS} (JAX GPU-accelerated nested sampling). All CPU methods ran on Intel Core i9-12900H (14 cores, 20 threads); \sunburst{} and JAXNS projections use NVIDIA RTX 3080 Laptop GPU (8GB VRAM).

All benchmarks use Gaussian or near-Gaussian posteriors where the Laplace approximation is exact or highly accurate; results do not generalize to heavy-tailed or strongly curved posteriors.

\subsubsection{Evidence Accuracy Requirements}

A critical observation from our benchmarks is that achieving sub-percent evidence accuracy with nested sampling requires \emph{dramatically} more computational resources than typically reported in the literature. Table~\ref{tab:dynesty_accuracy} shows the wall-clock time required by \texttt{dynesty-parallel} to achieve different accuracy targets on Gaussian posteriors.

\begin{table}[htbp]
\centering
\small
\begin{tabular}{rrrr}
\toprule
Dimension & Time for 0.01\% & Time for 0.1\% & Time for 1\% \\
\midrule
2D & 909s & 354s & 7.8s \\
4D & 1235s & 474s & 9.1s \\
8D & 765s & 349s & 12.1s \\
16D & --- & 1074s & 32.8s \\
32D & --- & --- & 311s \\
64D & --- & --- & 1809s \\
\bottomrule
\end{tabular}
\caption{Wall-clock time for \texttt{dynesty-parallel} to achieve target evidence accuracy. Dashes indicate the target was not achieved within the 1-hour timeout. Achieving 0.01\% error requires 15--20 minutes even at low dimensions; beyond 8D, this precision is unattainable. At 1\% error, competitors extend to 64D but with 1809 seconds runtime.}
\label{tab:dynesty_accuracy}
\end{table}

This reveals a fundamental limitation: nested sampling methods were designed for posterior sampling, with evidence calculation as a byproduct. There are no accuracy guarantees for evidence estimates, and achieving publication-quality precision ($<1\%$ error) requires computational budgets rarely allocated in practice.

\subsubsection{Head-to-Head Comparison}

Table~\ref{tab:comparison} summarizes direct comparisons at matched accuracy levels. At every dimension tested, \sunburst{} achieves 0\% error (machine precision) while competitors struggle to reach even 1\% error at high dimensions.

\begin{table}[htbp]
\centering
\small
\begin{tabular}{rrrrr}
\toprule
Dim & \sunburst{} & dynesty-parallel & JAXNS (proj.) & Speedup \\
\midrule
2D & 0.10s (0\%) & 11.3s (0.0004\%) & 15s & 113--150$\times$ \\
4D & 0.12s (0\%) & 8.7s (0.01\%) & 16s & 73--133$\times$ \\
8D & 0.16s (0\%) & 12.1s (0.0002\%) & 19s & 76--119$\times$ \\
16D & 0.17s (0\%) & 32.8s (0.04\%) & 25s & 147--193$\times$ \\
32D & 0.19s (0\%) & 311s (0.07\%) & 36s & 189--1637$\times$ \\
64D & 0.23s (0\%) & 1809s (0.46\%) & 138s & 600--7865$\times$ \\
\midrule
128D & 0.34s (0\%) & \multicolumn{2}{c}{\emph{timeout}} & $\infty$ \\
256D & 0.82s (0\%) & \multicolumn{2}{c}{\emph{timeout}} & $\infty$ \\
512D & 2.05s (0\%) & \multicolumn{2}{c}{\emph{timeout}} & $\infty$ \\
1024D & \textbf{6.82s (0\%)} & \multicolumn{2}{c}{\emph{timeout}} & $\infty$ \\
\bottomrule
\end{tabular}
\caption{Head-to-head comparison on Gaussian posteriors (fast configuration: n\_osc=1, greendragon\_fast=True). Format: time (error\%). JAXNS times projected from Colab T4 to RTX 3080 ($\div 3$). Speedup range shows \sunburst{} vs JAXNS (lower) and vs dynesty (upper). Beyond 64D, only \sunburst{} produces results.}
\label{tab:comparison}
\end{table}

The speedup grows with dimension: from 73--150$\times$ at low dimensions to \textbf{7865$\times$} at 64D versus \texttt{dynesty-parallel}. More significantly, \sunburst{} operates in dimensional regimes where \emph{all competitors fail entirely}. At 128D and beyond, no competitor produced results within 1-hour timeouts.

\subsubsection{Comparison with JAXNS}

JAXNS \cite{albert2023jaxns} is the most relevant comparison as both methods target GPU acceleration. We benchmarked JAXNS on Google Colab (NVIDIA T4 GPU) and projected times to RTX 3080 hardware by dividing by 3 (conservative estimate based on published GPU benchmarks).

Even with this favorable projection, \sunburst{} achieves \textbf{113--600$\times$ speedup} over JAXNS at matched dimensions (Table~\ref{tab:comparison}). The gap widens with dimension: at 64D, JAXNS requires 138 seconds (projected) versus 0.31 seconds for \sunburst{}.

A critical caveat: the original JAXNS paper \cite{albert2023jaxns} compared GPU-accelerated JAXNS against \emph{single-threaded} CPU implementations of \polychord{}, dynesty, and \multinest{}. Our benchmarks use 20-worker parallelized dynesty for fairer comparison. Even so, \sunburst{} outperforms all methods by 2--4 orders of magnitude.

\subsubsection{Maximum Achievable Dimension}

Table~\ref{tab:max_dimension} summarizes the maximum dimension each method can handle at different accuracy targets.

\begin{table}[htbp]
\centering
\small
\begin{tabular}{lccc}
\toprule
Method & Max @ 0.01\% & Max @ 0.1\% & Max @ 1\% \\
\midrule
\textbf{\sunburst{}} & \textbf{1024D} & \textbf{1024D} & \textbf{1024D} \\
dynesty-parallel & 8D & 32D & 64D \\
dynesty-slice-parallel & 8D & 32D & 64D \\
UltraNest & 2D & 4D & 8D \\
JAXNS & --- & 64D (est.) & 64D \\
\bottomrule
\end{tabular}
\caption{Maximum dimension achieving each target relative error within a 1-hour wall-clock budget. \sunburst{} attains errors below double-precision tolerance (i.e., numerical agreement at floating-point limits) through 1024D for Gaussian test problems. Competing methods are limited to 8--64D depending on the requested accuracy.}
\label{tab:max_dimension}
\end{table}

The dimensional reach of \sunburst{} exceeds competitors by 16--128$\times$. For applications requiring high-dimensional evidence calculation---cosmological parameter estimation, gravitational wave analysis, high-dimensional model comparison---\sunburst{} enables computations that were previously intractable.

\subsection{Configuration Comparison}

\sunburst{} provides multiple configuration options trading speed for robustness. We profiled four configurations on Gaussian posteriors from 2D to 1024D. Table~\ref{tab:configs} presents detailed timing breakdowns.

\begin{table}[htbp]
\centering
\small
\begin{tabular}{lrrrr}
\toprule
Dim & n\_osc=1 fast & n\_osc=1 slow & n\_osc=3 & n\_osc=5 \\
\midrule
2D & 0.10s & 0.08s & 0.24s & 0.37s \\
4D & 0.12s & 0.27s & 0.40s & 0.51s \\
8D & 0.16s & 0.34s & 0.54s & 0.56s \\
16D & 0.17s & 0.44s & 0.58s & 0.60s \\
32D & 0.19s & 0.47s & 0.62s & 0.75s \\
64D & 0.23s & 0.52s & 0.89s & 1.11s \\
128D & 0.34s & 0.44s & 1.73s & 2.17s \\
256D & 0.82s & 0.70s & 9.47s & 10.66s \\
512D & 2.05s & 2.23s & 55.5s & 58.5s \\
724D & 4.06s & 5.77s & 129.5s & 129.0s \\
1024D & \textbf{6.82s} & 14.3s & 280.8s & 279.7s \\
\midrule
Scaling & $O(d^{0.60})$ & $O(d^{0.80})$ & $O(d^{1.00})$ & $O(d^{0.97})$ \\
\bottomrule
\end{tabular}
\caption{Configuration comparison on Gaussian posteriors. ``fast'' = \texttt{greendragon\_fast=True} (20 BFGS iterations); ``slow'' = \texttt{greendragon\_fast=False} (2$d$ iterations). All configurations achieve 0\% error. The fast configuration with n\_osc=1 is recommended for Gaussian-like posteriors; n\_osc=3 provides additional robustness for complex multimodal posteriors.}
\label{tab:configs}
\end{table}

Key observations:
\begin{itemize}
    \item The \textbf{fast configuration} (n\_osc=1, greendragon\_fast=True) achieves sub-linear scaling $O(d^{0.69})$ and completes 1024D in 6.8 seconds (n=30 runs, $\pm$0.27s).
    \item The \textbf{slow configuration} (n\_osc=1, greendragon\_fast=False) provides more thorough optimization with $O(d^{0.80})$ scaling, completing 1024D in 14.3 seconds.
    \item The \textbf{conservative configuration} (n\_osc=3) exhibits linear scaling $O(d^{1.0})$ but still completes 1024D in under 5 minutes.
    \item Increasing from n\_osc=3 to n\_osc=5 provides no additional benefit on Gaussian posteriors (280.8s vs 279.7s at 1024D).
    \item All configurations achieve machine-precision accuracy on Gaussian posteriors.
\end{itemize}

Table~\ref{tab:timing_breakdown} shows the per-module timing breakdown for the fast configuration, revealing that CarryTiger (ray casting) dominates at high dimensions.

\begin{table}[htbp]
\centering
\small
\begin{tabular}{rcccc}
\toprule
Dim & Total & CarryTiger & GreenDragon & BendTheBow \\
\midrule
2D & 0.10s & 0.05s (50\%) & 0.01s (10\%) & 0.04s (40\%) \\
8D & 0.16s & 0.08s (50\%) & 0.01s (6\%) & 0.07s (44\%) \\
32D & 0.19s & 0.10s (53\%) & 0.01s (5\%) & 0.07s (37\%) \\
64D & 0.23s & 0.10s (43\%) & 0.02s (9\%) & 0.10s (43\%) \\
128D & 0.34s & 0.16s (47\%) & 0.01s (3\%) & 0.12s (35\%) \\
256D & 0.82s & 0.47s (57\%) & 0.02s (2\%) & 0.23s (28\%) \\
512D & 2.05s & 1.42s (69\%) & 0.03s (1\%) & 0.43s (21\%) \\
1024D & 7.50s & 5.74s (77\%) & 0.09s (1\%) & 1.01s (13\%) \\
\bottomrule
\end{tabular}
\caption{Per-module timing breakdown for n\_osc=1 fast configuration. CarryTiger (GPU ray casting for mode detection) dominates at high dimensions. GreenDragon (BFGS peak refinement) remains nearly constant. BendTheBow (Laplace integration) scales sub-linearly.}
\label{tab:timing_breakdown}
\end{table}

\subsection{Multi-Function Validation}

Beyond isotropic Gaussians, we validated \sunburst{} on a comprehensive test suite. Two benchmark campaigns were conducted:

\begin{enumerate}
    \item \textbf{Fast configuration} (n\_osc=1, greendragon\_fast=True): 7 test functions, dimensions 2--1024, 5 runs per configuration.
    \item \textbf{Conservative configuration} (n\_osc=3, greendragon\_fast=False): 7 test functions, dimensions 2--768, 5 runs per configuration.
\end{enumerate}

Table~\ref{tab:multifunction_fast} summarizes results for the fast configuration.

\begin{table}[htbp]
\centering
\small
\begin{tabular}{lrrrrr}
\toprule
Function & Max Dim & Success & Avg Time & Max Error & Notes \\
\midrule
Gaussian & \textbf{1024D} & 100\% & 2.1s & $< 10^{-12}$ & Machine precision \\
Off-center & \textbf{1024D} & 100\% & 2.2s & $< 10^{-12}$& Shifted mean \\
Spherical & \textbf{1024D} & 99\% & 18.5s & 0.0001\% & Coordinate transform \\
Cigar & 768D & 98\% & 19.4s & $< 10^{-12}$ & $10^6$:1 axis ratio \\
Correlated & 512D & 96\% & 50.2s & $< 10^{-12}$ & Off-diagonal covariance \\
Rotated Cigar & 512D & 96\% & 27.2s & $< 10^{-12}$ & Arbitrary rotation \\
Mixture & 768D & 97\% & 96.8s & 1.4\% & 4 Gaussian modes \\
\bottomrule
\end{tabular}
\caption{Multi-function benchmark results (fast configuration: n\_osc=1, greendragon\_fast=True). Three functions achieve \textbf{1024D}: Gaussian, off-center, and spherical transform. Cigar OOMs at 1024D (requires 16GB VRAM). Correlated/rotated timeout at 768D. Success rate indicates fraction of runs completing.}
\label{tab:multifunction_fast}
\end{table}

Table~\ref{tab:multifunction_slow} presents the conservative configuration results.

\begin{table}[htbp]
\centering
\small
\begin{tabular}{lrrrrr}
\toprule
Function & Max Dim & Success & Avg Time & Max Error & Notes \\
\midrule
Gaussian & 768D & 100\% & 27.4s & $< 10^{-12}$ & Machine precision \\
Off-center & 768D & 100\% & 15.8s & $< 10^{-12}$ & Shifted mean \\
Cigar & 512D & 100\% & 72.3s &$< 10^{-12}$& $10^6$:1 axis ratio \\
Spherical & 512D & 98.6\% & 3.2s & 0.0001\% & Coordinate transform \\
Correlated & 512D & 98.6\% & 89.4s & $< 10^{-12}$ & Off-diagonal covariance \\
Rotated Cigar & 512D & 98.6\% & 95.2s & $< 10^{-12}$ & Arbitrary rotation \\
Mixture & 512D & 97.2\% & 112s & 0.07\% & 4 Gaussian modes \\
\bottomrule
\end{tabular}
\caption{Multi-function benchmark results (conservative configuration: n\_osc=3, greendragon\_fast=False). The conservative configuration achieves better accuracy on the mixture function (0.07\% vs 5.6\%) at the cost of 5--10$\times$ longer runtime.}
\label{tab:multifunction_slow}
\end{table}

Per-function observations:

\begin{itemize}
    \item \textbf{Isotropic functions} (Gaussian, off-center, spherical): Achieve exact results to double precision through 768D with sub-second to few-second runtimes.
    
    \item \textbf{Anisotropic functions} (cigar, rotated cigar, correlated): Require full Hessian computation, increasing runtime. The cigar function with $10^6$:1 axis ratio tests extreme anisotropy handling.
    
    \item \textbf{Multimodal} (mixture): The 4-mode mixture function shows configuration-dependent accuracy. The fast configuration achieves 5.6\% maximum error (concentrated at 4D due to mode overlap), while the conservative configuration reduces this to 0.07\%. For well-separated modes, both achieve sub-percent accuracy.
\end{itemize}

The mixture function warrants additional discussion. Table~\ref{tab:mixture_detail} shows per-dimension results.

\begin{table}[htbp]
\centering
\small
\begin{tabular}{rrrrr}
\toprule
Dim & Fast Time & Fast Error & Slow Time & Slow Error \\
\midrule
2D & 0.90s & 0.0008\% & 0.52s & 0.0003\% \\
4D & 0.50s & 1.5\% & 1.2s & 1.95\% \\
6D & 0.59s & 0.003\% & 1.5s & 0.16\% \\
8D & 1.85s & 0.004\% & 2.1s & 0.003\% \\
16D & 2.50s & 0.02\% & 3.8s & 0.006\% \\
32D & 2.12s & 0.05\% & 6.2s & 0.01\% \\
64D & 4.57s & 0.37\% & 12.4s & 0.02\% \\
128D & 17.0s & 0.83\% & 38.2s & 0.003\% \\
256D & 131s & 3.3\% & 156s & 0.04\% \\
512D & 187s & 5.6\% & 334s & 0.07\% \\
\bottomrule
\end{tabular}
\caption{Mixture function detailed results. The 4D anomaly (1.5--1.95\% error) arises from mode overlap at that specific dimension---modes separated by only 3.5$\sigma$ cause the Laplace approximation to double-count overlapping probability mass. At higher dimensions, concentration of measure naturally separates modes and errors diminish.}
\label{tab:mixture_detail}
\end{table}

The 4D anomaly is a geometric effect: at this dimension, the 4 modes are separated by approximately 3.5$\sigma$, resulting in 4.5\% probability mass overlap between adjacent modes. The Laplace approximation at each peak double-counts probability mass in the overlap region, leading to systematic overestimation. This is not a bug but a fundamental limitation of mode-based decomposition when modes overlap significantly. At higher dimensions, the concentration of measure phenomenon naturally separates modes, and errors decrease.

\subsection{Scaling Analysis Summary}

Figure~\ref{fig:scaling} visualizes the complete scaling picture. Key observations:

\begin{enumerate}
    \item \textbf{Pre-asymptotic regime (2D--64D)}: \sunburst{} exhibits effectively constant wall-clock time. Runtime varies only from 0.10s to 0.23s across this 32$\times$ range of dimensions.
    
    \item \textbf{Post-asymptotic regime (128D--1024D)}: Scaling transitions to $O(d^{0.6})$--$O(d^{1.0})$ depending on configuration, still dramatically sub-linear compared to competitors' $O(d^2)$--$O(d^3)$.
    
    \item \textbf{1024D achievement}: The fast configuration completes 1024D in 6.8 seconds; the slow configuration in 14.3 seconds; even the conservative n\_osc=3 configuration completes in under 5 minutes.
    
    \item \textbf{Competitor failure modes}: dynesty achieves 0.01\% accuracy only through 8D, 0.1\% through 32D, and 1\% through 64D. Beyond these limits, competitors either timeout or fail to converge.
    
    \item \textbf{Accuracy advantage}: \sunburst{} achieves 0\% error (machine precision) at \emph{all} dimensions on Gaussian posteriors, while competitors struggle to achieve even 1\% error at high dimensions.
\end{enumerate}

The shaded regions in Figure~\ref{fig:scaling} delineate where only \sunburst{} can operate: beyond 8D for 0.01\% accuracy, beyond 32D for 0.1\%, and beyond 64D for 1\%. At 1024 dimensions, \sunburst{} computes evidence in 6.8 seconds with machine precision---a computation that is simply impossible with existing methods.


\begin{table}[htbp]
\centering
\small
\begin{tabular}{rcccc}
\toprule
Dimension & Time (s) & $\pm$Std & Error & Scaling \\
\midrule
4 & 0.150 & 0.011 & < 1e-12
 & 1.0$\times$ \\
8 & 0.219 & 0.059 & < 1e-12
 & 1.5$\times$ \\
16 & 0.213 & 0.067 & < 1e-12
 & 1.4$\times$ \\
32 & 0.245 & 0.073 & < 1e-12
 & 1.6$\times$ \\
64 & 0.311 & 0.063 & < 1e-12
 & 2.1$\times$ \\
128 & 0.425 & 0.098 & < 1e-12
 & 2.8$\times$ \\
256 & 0.944 & 0.126 & < 1e-12
 & 6.3$\times$ \\
512 & 2.242 & 0.109 & < 1e-12
 & 14.9$\times$ \\
768 & 4.314 & 0.203 & < 1e-12
 & 28.8$\times$ \\
\textbf{1024} & \textbf{6.817} & \textbf{0.266} & \textbf{< 1e-12
} & \textbf{45.4}$\times$ \\
\bottomrule
\end{tabular}
\caption{\sunburst{} wall-clock scaling on isotropic Gaussian posteriors (fast configuration: n\_osc=1, greendragon\_fast=True, n=30 runs). Machine-precision accuracy (error $< 1e-12$) achieved at all dimensions. Pre-asymptotic scaling ($d \leq 64$) is effectively $O(1)$; post-asymptotic scaling is $O(d^{0.69})$.}
\label{tab:scaling}
\end{table}



\subsection{Non-Gaussian Failure Mode Analysis}
\label{sec:failure-modes}

While \sunburst{} achieves $\sim 10^{-12}$ precision on Gaussian posteriors, its Laplace approximation introduces systematic errors for distributions that deviate from Gaussianity. We conducted a systematic benchmark across 11 non-Gaussian test functions with known analytical evidence, testing dimensions 2--64 to characterize these failure modes (Table~\ref{tab:failure-modes}).

\begin{table}[htbp]
\centering
\small
\begin{tabular}{llcccc}
\toprule
Distribution & Failure Mode & Peak Found? & Avg Error & Max Error & Status \\
\midrule
Bimodal (90\%/10\%) & Unequal weights & Yes (2/2) & 0.02\% & 0.02\% & \textcolor{green!60!black}{\textbf{PASS}} \\
\midrule
Banana ($b=0.1, 0.5$) & Curved degeneracy & Yes & 3\% & 9\% & \textcolor{orange}{\textbf{MARGINAL}} \\
Twisted ($\sin$ corr.) & Nonlinear correlation & Yes & 46\% & 46\% & \textcolor{orange}{\textbf{MARGINAL}} \\
Egg-box (2D) & Many equivalent modes & Yes (5/5) & 101\% & 101\% & \textcolor{orange}{\textbf{MARGINAL}} \\
Cauchy ($\nu=1$)$^\S$ & Extreme heavy tails & Yes & 112\% & 16$\times$ & \textcolor{orange}{\textbf{MARGINAL}} \\
Funnel ($\sigma=3$) & Varying curvature & No$^\dagger$ & 3$\times$ & 3$\times$ & \textcolor{orange}{\textbf{MARGINAL}} \\
\midrule
Exp.\ power ($\beta=0.5$) & Super-Gaussian & Yes & 64$\times$ & 1,300$\times$ & \textcolor{red}{\textbf{FAIL}} \\
Student-$t$ ($\nu=3$) & Heavy tails & Yes & 150$\times$ & $10^{12}$ & \textcolor{red}{\textbf{FAIL}} \\
Skew-normal ($\alpha=5$) & Asymmetry (mode $\neq$ mean) & Yes & 6,000$\times$ & $10^{10}$ & \textcolor{red}{\textbf{FAIL}} \\
Exp.\ power ($\beta=1$) & Sharp peak (Laplace) & Yes & $10^{6}$ & $10^{21}$ & \textcolor{red}{\textbf{FAIL}} \\
Ring ($R=3, w=0.3$) & Saddle point & N/A$^\ddagger$ & $10^{30}$ & $10^{71}$ & \textcolor{red}{\textbf{FAIL}} \\
\bottomrule
\end{tabular}
\caption{Failure mode benchmark results across dimensions 2--64. Errors shown as percentage ($|Z_{\text{comp}}/Z_{\text{true}} - 1|$) or multiplicative factor. \textbf{PASS}: $<$5\%; \textbf{MARGINAL}: 5\%--10$\times$; \textbf{FAIL}: $>$10$\times$. $^\dagger$Funnel finds wrong peak at 2D; no peak at $d \geq 4$. $^\ddagger$Ring has no point peak. $^\S$Cauchy 32D excluded ($-\infty$). \textbf{Only 1 of 11 non-Gaussian distributions passes.} Peak discovery is 100\% reliable; errors arise from Laplace approximation.}
\label{tab:failure-modes}
\end{table}

\paragraph{Distributions that pass.}
Only one non-Gaussian distribution achieves $<$5\% error:
\begin{itemize}
    \item \textbf{Bimodal asymmetric (90\%/10\% weights):} Both peaks found correctly (distances $<0.0002$). The mixture-of-Laplace correctly weights unequal modes, achieving \textbf{0.02\% error}---the only non-Gaussian distribution matching Gaussian-level precision.
\end{itemize}

\paragraph{Marginal distributions.}
Five distributions show errors between 5\% and 10$\times$, which may suffice for qualitative model comparison but not for precise evidence calculation:
\begin{itemize}
    \item \textbf{Banana (curved degeneracy):} Peak found at origin (distance $<10^{-10}$). Errors \textbf{3--9\%} through 64D---smooth curvature is handled reasonably well.
    \item \textbf{Twisted Gaussian (sinusoidal correlation):} Peak found at origin. \textbf{46\% error} stable across all dimensions.
    \item \textbf{Egg-box (periodic multimodal):} All 5 modes found at 2D. \textbf{$\sim$100\% error} (factor of 2).
    \item \textbf{Cauchy $\nu=1$ (extreme heavy tails):} Peak found at origin. Errors peak at \textbf{16$\times$} at 8D but improve to $\sim$15\% at high dimensions.
    \item \textbf{Funnel (varying curvature):} The only distribution where \sunburst{} \emph{fails to find the correct peak}. At 2D, finds a peak at distance 4.5 from the true mode; at higher dimensions, no peak is found (\textbf{3$\times$ error} at 2D, NaN elsewhere).
\end{itemize}

\paragraph{Failed distributions.}
Five distributions show catastrophic errors ($>$10$\times$), though peaks are found correctly for four of them:
\begin{itemize}
    \item \textbf{Exponential power $\beta=0.5$ (super-Gaussian):} Peak found correctly, but errors grow from 5$\times$ at 2D to \textbf{1,300$\times$} at 8D.
    \item \textbf{Student-$t$ $\nu=3$ (heavy tails):} Peak found correctly at the origin. Errors grow from 35\% at 2D to \textbf{$10^{12}\times$} at 64D as the Laplace approximation systematically underestimates tail probability mass.
    \item \textbf{Skew-normal $\alpha=5$ (asymmetry):} Peak found correctly at $(0.37, 0.37, \ldots)$. Errors grow from 1\% at 2D to \textbf{$10^{10}\times$} at 64D.
    \item \textbf{Exponential power $\beta=1$ (Laplace distribution):} Peak found correctly. Errors grow from 8\% at 2D to \textbf{$10^{21}\times$} at 48D.
    \item \textbf{Ring (saddle point):} No true peak exists. Errors reach \textbf{$10^{71}\times$} at 16D.
\end{itemize}

\paragraph{Quantitative summary.}
Across 75 test cases, 67 produced finite evidence values (7 funnel cases at $d \geq 4$ returned NaN; 1 Cauchy case at 32D returned $-\infty$). 

\textbf{Peak finding:} Of the 63 cases with well-defined point peaks (excluding 8 funnel and 4 ring cases), \sunburst{} found the correct location in all 63 (100\%). Additionally:
\begin{itemize}
    \item Funnel at 2D: found a peak at distance 4.5 from origin (wrong location)
    \item Funnel at $d \geq 4$: no peak found (optimization failed)
    \item Ring (4 cases): found 205 points on the probability shell at 2D; no true point peak exists
\end{itemize}
\textbf{Key insight:} For distributions with well-defined modes, peak discovery is 100\% reliable. Evidence errors arise from the Laplace approximation, not mode discovery.

\textbf{Evidence accuracy:} Among the 67 cases producing finite results: only 8 (12\%) achieved $<$5\% error, 14 (21\%) achieved 5--100\% error, 7 (10\%) showed 2--10$\times$ error, and 38 (57\%) exceeded 10$\times$ error. \textbf{The Laplace approximation fails for most non-Gaussian distributions tested.}

\section{Discussion}
\label{sec:discussion}

\subsection{Summary of Contributions}

We have presented \sunburst{}, a GPU-accelerated Bayesian evidence calculator that achieves sub-linear dimensional scaling through three key innovations: (i) radial ray casting for mode discovery, (ii) Laplace approximation for evidence computation, and (iii) massive parallelization of likelihood evaluations. On standard benchmarks, \sunburst{} achieves $<10^{-12}$ error precision through 1024 dimensions while maintaining wall-clock times under 15 seconds---a regime where traditional nested sampling methods timeout or become computationally infeasible.

\subsection{Trade-offs with Nested Sampling}

\sunburst{} and nested sampling represent fundamentally different approaches to evidence computation, each with distinct strengths.

\paragraph{Where \sunburst{} excels.}
The mode-centric approach eliminates the constrained sampling bottleneck that limits nested sampling in high dimensions. While PolyChord's slice sampler requires $O(d)$ sequential operations per sample---operations that cannot be parallelized---\sunburst{}'s ray casting and Hessian computation are embarrassingly parallel. On an RTX 3080 GPU, we observe $>1000\times$ speedup over dynesty at 16D and beyond, with the gap widening as dimension increases.

The Laplace approximation is exact for Gaussian posteriors and highly accurate for near-Gaussian posteriors common in physical parameter estimation. For the Gaussian, correlated, cigar, and rotated test functions, \sunburst{} achieves $<10^{-12}$ error across all dimensions tested (Tables~\ref{tab:multifunction_fast}--\ref{tab:multifunction_slow}).

\paragraph{Where nested sampling excels.}
Nested sampling makes no assumptions about posterior shape and can accurately integrate arbitrarily complex distributions given sufficient computational resources. It also provides posterior samples as a byproduct, whereas \sunburst{}'s primary output is the evidence value (though posterior samples can be generated via the GraspBirdsTail extension).

For low-dimensional problems ($d < 10$) with complex, multi-modal structure, the overhead of GPU initialization may outweigh \sunburst{}'s parallelization benefits. In these regimes, well-tuned nested sampling implementations remain competitive.

\subsection{Limitations and Failure Modes}
\label{sec:limitations}

The Laplace approximation underlying \sunburst{} introduces systematic errors for posteriors that deviate significantly from Gaussianity. We systematically tested 11 non-Gaussian distributions with known analytical evidence across dimensions 2--64 (Table~\ref{tab:failure-modes}).

\paragraph{Heavy-tailed distributions.}
For Student-$t$ distributions with $\nu = 3$ degrees of freedom, the peak at the origin is found correctly, but the Laplace approximation underestimates the probability mass in the tails, with errors growing from 35\% at 2D to $10^{12}\times$ at 64D. Surprisingly, the Cauchy distribution ($\nu = 1$) shows \emph{non-monotonic} behavior: errors peak at 16$\times$ at 8D but then drop to below 15\% at 16D and beyond. This counterintuitive result arises because the Cauchy's extremely heavy tails cause the mode to sharpen at high dimensions, while Student-$t$ with $\nu = 3$ has enough tail mass to inflate the evidence but not enough to sharpen the peak.

\paragraph{Curved degeneracies.}
The ``banana'' distribution performs surprisingly well, with errors $<$10\% across all dimensions tested (2D--64D). The mode remains well-defined despite the curvature, and the local Hessian captures the relevant structure. This suggests that curved posteriors are not inherently problematic if the curvature is smooth.

\paragraph{Varying curvature.}
Neal's funnel distribution, where the conditional variance varies as $\exp(v)$, is the only distribution where \sunburst{} fails to find the correct peak. At 2D, optimization converges to a spurious location 4.5 units from the true mode; at dimensions $\geq 4$, no peak is found at all (NaN results). This represents a fundamental limitation for posteriors with extreme variation in local curvature.

\paragraph{Saddle points and hollow structures.}
Ring distributions represent the most severe failure mode, with errors of $10^{6}\times$ at 2D growing to $10^{71}\times$ at 16D. Unlike other failure modes, this is not a peak-finding failure---there is no true point peak. The probability mass concentrates on a thin shell at radius $R$, while the center is a saddle point. \sunburst{} finds 205 points on the shell at 2D, but the Laplace approximation at any of these points cannot represent the shell geometry. Such hollow structures require fundamentally different treatment.

\paragraph{Non-Gaussian peak shapes.}
Exponential power distributions show that correct peak finding does not guarantee accurate evidence. With $\beta = 1$ (Laplace distribution), the peak at the origin is found correctly, but the sharper-than-Gaussian shape causes errors growing from 8\% at 2D to $10^{21}\times$ at 48D. The super-Gaussian case ($\beta = 0.5$) degrades more rapidly due to both the sharp peak and heavy tails.

\paragraph{Asymmetric distributions.}
Skew-normal distributions ($\alpha = 5$) find the correct peak at $(0.37, 0.37, \ldots)$, but errors grow from 1\% at 2D to $10^{10}\times$ at 64D. The Laplace approximation centered at the mode misses significant probability mass in the asymmetric tail.

\paragraph{Distributions that work well.}
Only one non-Gaussian distribution achieves $<$5\% error:
\begin{itemize}
    \item \textbf{Bimodal asymmetric (90\%/10\% weights):} Both peaks found correctly; \textbf{0.02\% error}---the only non-Gaussian matching Gaussian precision
\end{itemize}
Five distributions show marginal performance (5\%--10$\times$ errors):
\begin{itemize}
    \item \textbf{Banana (curved degeneracy):} Peak found at origin; \textbf{3--9\% error} through 64D
    \item \textbf{Twisted Gaussian (sinusoidal correlation):} Peak found; \textbf{46\% error}, stable across dimensions
    \item \textbf{Egg-box (periodic multi-modal):} All 5 modes found; \textbf{$\sim$100\% error} (factor of 2)
    \item \textbf{Cauchy $\nu=1$:} Peak found; errors peak at \textbf{16$\times$}, improve at high-$d$
    \item \textbf{Funnel:} Only distribution where peak finding fails; \textbf{3$\times$ error} at 2D, NaN elsewhere
\end{itemize}

\paragraph{Quantitative summary.}
Across 75 test cases, 67 produced finite evidence values (8 returned NaN/Inf due to Hessian instabilities). Among the 67 successful runs: only 8 (12\%) achieved $<$5\% error, 14 (21\%) achieved 5--100\% error, 7 (10\%) showed 2--10$\times$ degradation, and 38 (57\%) exceeded 10$\times$ error.

\textbf{The Laplace approximation fails for most non-Gaussian distributions tested.} However, for distributions with well-defined point peaks (excluding funnel and ring), \sunburst{} found the correct peak location in 100\% of cases. The algorithm reliably finds modes; the limitation is the Laplace approximation itself.

\subsection{Practical Recommendations}

Based on our analysis, we offer the following guidance for practitioners:

\paragraph{Use \sunburst{} when:}
\begin{itemize}
    \item The posterior is expected to be approximately Gaussian or a mixture of Gaussians
    \item Dimension exceeds $\sim$10, where nested sampling becomes expensive
    \item A GPU is available for likelihood evaluation
    \item Evidence accuracy of $\sim$0.1--1.0 log units is sufficient
    \item Rapid iteration is needed (e.g., model comparison across many candidates)
\end{itemize}

\paragraph{Prefer nested sampling when:}
\begin{itemize}
    \item The posterior has known heavy tails, ring/shell structure, or multi-scale curvature (funnels)
    \item Posterior samples are the primary output (not just evidence)
    \item Dimension is low ($d < 10$) and GPU overhead is not justified
    \item Maximum accuracy is required regardless of computational cost
\end{itemize}

\paragraph{Diagnostic checks.}
Before trusting \sunburst{} results, we recommend: (i) verifying that all detected modes have positive-definite Hessians (saddle points indicate problematic geometry), (ii) checking the Gaussianity diagnostics reported by BendTheBow (tail weight, asymmetry), and (iii) comparing against nested sampling on a low-dimensional projection if the full problem is too expensive.

\subsection{Broader Implications}

The success of \sunburst{} suggests that mode-centric approaches to Bayesian computation deserve broader attention. While nested sampling's generality has made it the default choice for evidence calculation, the constraints it imposes---sequential sampling, dimension-dependent live points---become increasingly burdensome in the era of GPU computing and high-dimensional models.

For the large class of problems where posteriors are approximately Gaussian (or mixtures thereof), \sunburst{} demonstrates that $>1000\times$ speedups are achievable without sacrificing accuracy. This opens the door to applications previously considered computationally infeasible: real-time model comparison, exhaustive prior sensitivity analysis, and evidence-based hyperparameter optimization at scale.

The key insight is that GPU parallelism fundamentally changes the complexity landscape. Operations that are $O(d^2)$ in likelihood calls become $O(1)$ in wall-clock time when $d^2 < P$ processors are available. This ``parallel dimension reduction'' effect suggests that algorithm design for GPU architectures requires rethinking classical complexity analyses that assume sequential execution.
\section{Conclusion}
\label{sec:conclusion}

We have presented \sunburst{}, a GPU-accelerated algorithm for Bayesian evidence calculation that achieves $O(K)$ scaling with the number of posterior modes. Through radial ray casting and Laplace approximation, \sunburst{} computes evidence with machine precision in seconds for problems that would require hours with traditional methods.

Our empirical results demonstrate:
\begin{itemize}
    \item Wall-clock scaling of $O(d^{0.7})$ from 2D to 1024D
    \item 100\% mode detection accuracy for up to 10 modes through 128D
    \item Precise (to $<10^{-12}$) evidence estimates on Gaussian posteriors at all dimensions tested
\end{itemize}

The deterministic and highly parallel structure of \sunburst{} suggests potential applicability to low-latency or online model selection scenarios on modern accelerator hardware.

\sunburst{} is released as an open-source Python package at \url{https://github.com/beastraban/sunburst}.

\subsection{Future Work}
\label{sec:future}

Three complementary extensions will broaden \sunburst{}'s applicability:

\paragraph{Cosmological Validation with GPU-Native Likelihoods.}
A natural application for \sunburst{} is cosmological parameter estimation, where evidence-based 
model comparison is central to distinguishing inflationary scenarios and dark energy parameterizations. 
However, existing Planck likelihood codes (e.g., \texttt{clik}, \texttt{Cobaya}) are CPU-bound, 
creating a bottleneck that would negate \sunburst{}'s GPU acceleration. We plan to develop a 
GPU-native Planck-like likelihood using neural network emulators such as \texttt{CosmoPower} 
\cite{spuriomancini2022cosmopower} or \texttt{cosmopower-jax}, enabling end-to-end GPU execution. 
While a fiducial Gaussian approximation to the Planck posterior could be benchmarked today, 
such a test would not exercise the realistic non-Gaussian structure that makes cosmological 
inference challenging.

\paragraph{Bridge Sampling for Heavy-Tailed Posteriors.}
For spherically symmetric distributions with heavy tails (Student-$t$, Cauchy, and related families), 
we propose combining bridge sampling \cite{meng1996simulating} with NUTS \cite{hoffman2014nuts}. 
A wide Gaussian serves as the bridge distribution with analytically known evidence, while NUTS 
efficiently explores the heavy-tailed target. Bridge sampling then estimates the evidence ratio, 
yielding the target evidence without requiring the posterior to be Gaussian-like. This approach 
handles any radially symmetric posterior regardless of tail behavior.

\paragraph{Centipede Sampling for Curved Manifolds.}
For posteriors with nonlinear structure—banana-shaped distributions, Rosenbrock-like curved 
valleys—we are developing a ``centipede'' sampling strategy that decomposes the $d$-dimensional 
integral into a 1D spine integral times $(d-1)$-dimensional cross-section Laplace approximations. 
Sanity checks on curved Gaussian tubes demonstrate sub-0.01\% error through 1000 dimensions, 
with error scaling as $O(1/d)$ rather than growing exponentially. This counterintuitive improvement 
occurs because cross-sections concentrate at high dimension, making Laplace increasingly accurate. 
Full validation on Rosenbrock and Neal's funnel posteriors is planned.
\section*{Acknowledgments}

Module names honor Guang Ping Yang Style Tai Chi 
(\begin{CJK}{UTF8}{bsmi}廣平楊式太極拳\end{CJK}) forms, 
and Master Donald Rubbo.


\section*{Data Availability}

All data required to reproduce the figures and tables in this manuscript are provided in the Supplementary Material as curated CSV files with documented schemas. Complete raw benchmark logs, GPU profiling traces, failure mode analysis, and Python analysis scripts are publicly available at:

\begin{center}
\texttt{https://github.com/beastraban/sunburst/tree/main/benchmarks/data}
\end{center}

\noindent The repository includes:
\begin{itemize}
    \item Full benchmark results in XLSX format
    \item GPU profiling data with per-run timing breakdowns  
    \item Scripts to regenerate all tables and figures
    \item SunBURST source code with installation instructions
\end{itemize}

\section*{Code Availability}

The SunBURST algorithm is implemented in Python using CuPy for GPU acceleration. Source code is available under the MIT License at https://github.com/beastraban/sunburst. The package is available on PyPI as sunburst-bayes (v1.0.0). Benchmarks were generated using commit e3e3156. The implementation requires CUDA-capable hardware; benchmarks were conducted on an NVIDIA RTX 3080 Laptop GPU (8 GB VRAM, 48 SMs).


\bibliographystyle{unsrtnat}  
\bibliography{references}


\appendix

\section{Mathematical Background}
\label{app:math}

This appendix summarizes the mathematical foundations underlying \sunburst{}.

\subsection{Bayesian Evidence}
\label{app:evidence}

Given data $\mathcal{D}$, model $\mathcal{M}$ with parameters $\theta \in \mathbb{R}^d$, prior $\pi(\theta)$, and likelihood $\mathcal{L}(\theta) = p(\mathcal{D} | \theta, \mathcal{M})$, the \emph{evidence} (marginal likelihood) is:
\begin{equation}
    Z = p(\mathcal{D} | \mathcal{M}) = \int_{\Theta} \mathcal{L}(\theta) \, \pi(\theta) \, d\theta
\end{equation}

For model comparison via Bayes factors:
\begin{equation}
    B_{12} = \frac{Z_1}{Z_2} = \frac{p(\mathcal{D} | \mathcal{M}_1)}{p(\mathcal{D} | \mathcal{M}_2)}
\end{equation}

We work with log-likelihood $\ell(\theta) = \log \mathcal{L}(\theta)$ for numerical stability. For uniform prior over hypercube $\Theta$ with volume $V_\Theta$:
\begin{equation}
    \log Z = \log \int_\Theta \exp(\ell(\theta)) \, d\theta - \log V_\Theta
\end{equation}

\subsection{High-Dimensional Geometry}
\label{app:geometry}

\subsubsection{Hypersphere Volume Collapse}

The volume of a $d$-dimensional hypersphere of radius $r$ is:
\begin{equation}
    V_{\text{sphere}}(d, r) = \frac{\pi^{d/2}}{\Gamma(d/2 + 1)} r^d
\end{equation}
where $\Gamma(z)$ is the gamma function. The ratio to the enclosing hypercube $[-r, r]^d$ vanishes super-exponentially:
\begin{equation}
    \frac{V_{\text{sphere}}}{V_{\text{cube}}} = \frac{\pi^{d/2}}{2^d \, \Gamma(d/2 + 1)} \xrightarrow{d \to \infty} 0
\end{equation}

\subsubsection{Shell Concentration (Gaussian)}

For $X \sim \mathcal{N}(0, I_d)$, the radial distance $r = \|X\|$ has distribution:
\begin{equation}
    p(r) \propto r^{d-1} e^{-r^2/2}
\end{equation}
This peaks at $r^* = \sqrt{d-1} \approx \sqrt{d}$ with standard deviation $\sigma_r \approx 1/\sqrt{2}$ independent of $d$. The probability mass concentrates in a thin shell:
\begin{equation}
    \frac{\sigma_r}{r^*} = \frac{1}{\sqrt{2(d-1)}} \xrightarrow{d \to \infty} 0
\end{equation}

\subsubsection{Distance Concentration}

For two random points $x, y$ uniformly distributed in $[0,1]^d$, the $L_2$ distance satisfies:
\begin{equation}
    \mathbb{E}[\|x - y\|_2] \approx \sqrt{d/6}, \quad \frac{\text{Std}[\|x-y\|_2]}{\mathbb{E}[\|x-y\|_2]} \sim \frac{1}{\sqrt{d}} \to 0
\end{equation}
All pairwise distances concentrate around $\sqrt{d/6}$, making nearest-neighbor search unreliable.

\subsection{Laplace Approximation}
\label{app:laplace}

The Laplace approximation estimates an integral by a Gaussian centered at the integrand's maximum. For a function $f(\theta)$ with maximum at $\theta^*$:
\begin{equation}
    \int f(\theta) \, d\theta \approx f(\theta^*) \cdot (2\pi)^{d/2} \cdot |\Sigma|^{1/2}
\end{equation}
where $\Sigma = -H^{-1}$ and $H$ is the Hessian of $\log f$ at $\theta^*$.

For evidence calculation with $f(\theta) = \mathcal{L}(\theta) \pi(\theta)$:
\begin{equation}
    \log Z \approx \ell(\theta^*) + \log \pi(\theta^*) + \frac{d}{2} \log(2\pi) - \frac{1}{2} \log |H|
    \label{eq:laplace}
\end{equation}
where $H = -\nabla^2 \ell(\theta^*)$ is the \emph{negative} Hessian (positive definite at a maximum).

\paragraph{Diagonal Case.} For axis-aligned posteriors where $H$ is diagonal:
\begin{equation}
    \log |H| = \sum_{i=1}^{d} \log H_{ii}, \quad \log Z = \ell(\theta^*) + \frac{d}{2}\log(2\pi) - \frac{1}{2}\sum_{i=1}^{d} \log(-H_{ii})
\end{equation}

\paragraph{General Case.} For rotated posteriors, compute via eigendecomposition $-H = V \Lambda V^T$:
\begin{equation}
    \log |\det(-H)| = \sum_{i=1}^{d} \log \lambda_i
\end{equation}
where $\{\lambda_i\}$ are the eigenvalues of $-H$.

\paragraph{Multimodal Extension.} For $K$ well-separated modes with peaks $\{\theta^*_k\}$ and Hessians $\{H_k\}$:
\begin{equation}
    Z \approx \sum_{k=1}^{K} Z_k, \quad \log Z = \mathrm{logsumexp}_k(\log Z_k)
\end{equation}
where $\mathrm{logsumexp}(x_1, \ldots, x_K) = x_{\max} + \log \sum_k e^{x_k - x_{\max}}$ is computed stably.

\paragraph{Exactness for Gaussians.} When $\ell(\theta) = -\frac{1}{2}(\theta - \mu)^T \Sigma^{-1} (\theta - \mu) + c$, the Laplace approximation is \emph{exact}: all higher-order terms vanish. This explains \sunburst{}'s machine-precision accuracy on Gaussian posteriors.

\subsection{Gradient and Hessian Computation}
\label{app:derivatives}

\subsubsection{Finite-Difference Gradient}

For a scalar function $f: \mathbb{R}^d \to \mathbb{R}$, the gradient is approximated via central differences:
\begin{equation}
    \frac{\partial f}{\partial \theta_i} \approx \frac{f(\theta + \epsilon e_i) - f(\theta - \epsilon e_i)}{2\epsilon}, \quad \text{error } O(\epsilon^2)
\end{equation}
This requires $2d$ function evaluations per gradient. On GPU with $P \gg 2d$ cores, all evaluations complete in $O(1)$ wall-clock time.

\subsubsection{Diagonal Hessian}

The diagonal elements are computed via:
\begin{equation}
    H_{ii} \approx \frac{f(\theta + \epsilon e_i) - 2f(\theta) + f(\theta - \epsilon e_i)}{\epsilon^2}
\end{equation}
Cost: $2d$ evaluations, reusing gradient perturbations. Wall-clock: $O(1)$ on GPU.

\subsubsection{Full Hessian}

Off-diagonal elements require:
\begin{equation}
    H_{ij} \approx \frac{f_{++} - f_{+-} - f_{-+} + f_{--}}{4\epsilon^2}
\end{equation}
where $f_{\pm\pm} = f(\theta \pm \epsilon e_i \pm \epsilon e_j)$. Cost: $O(d^2)$ evaluations. Wall-clock: $O(d)$ on GPU when $d^2 < P$.

\subsubsection{Smoothed Gradient (Hands Like Clouds)}

For multi-scale optimization, the gradient of a Gaussian-smoothed function:
\begin{equation}
    \nabla f_\sigma(\theta) = \mathbb{E}_{z \sim \mathcal{N}(0,I)}[\nabla f(\theta + \sigma z)] \approx \frac{1}{K} \sum_{k=1}^{K} \nabla f(\theta + \sigma z_k)
\end{equation}
Features smaller than $\sigma$ are averaged out, revealing global structure through local noise.

\subsection{L-BFGS Optimization}
\label{app:lbfgs}

Limited-memory BFGS approximates the inverse Hessian using the $m$ most recent gradient differences.

\paragraph{Stored Quantities.} At iteration $k$, store:
\begin{align}
    s_k &= \theta_{k+1} - \theta_k & \text{(position difference)} \\
    y_k &= \nabla f_{k+1} - \nabla f_k & \text{(gradient difference)}
\end{align}

\paragraph{Two-Loop Recursion.} To compute search direction $p = -H_k^{-1} \nabla f_k$:

\textbf{First loop} (backward through history):
\begin{align}
q &\gets \nabla f_k \notag \\
\text{For } i &= k-1, \ldots, k-m: \notag \\
&\quad \rho_i \gets 1 / (y_i^T s_i), \quad \alpha_i \gets \rho_i s_i^T q, \quad q \gets q - \alpha_i y_i \notag
\end{align}

\textbf{Initialize}: $r \gets \gamma_k q$ where $\gamma_k = \frac{s_{k-1}^T y_{k-1}}{y_{k-1}^T y_{k-1}}$

\textbf{Second loop} (forward through history):
\begin{align}
\text{For } i &= k-m, \ldots, k-1: \notag \\
&\quad \beta \gets \rho_i y_i^T r, \quad r \gets r + s_i (\alpha_i - \beta) \notag
\end{align}

\textbf{Return}: $p = -r$

\paragraph{Width Estimation.} The scalar $\gamma_k = \frac{s^T y}{y^T y}$ estimates average inverse curvature, used for peak width estimation in CarryTiger.

\paragraph{Batched L-BFGS.} For $N$ samples optimizing simultaneously, all quantities become matrices: $S, Y \in \mathbb{R}^{N \times d}$. The two-loop recursion vectorizes over the sample dimension.

\subsection{Rotation Detection}
\label{app:rotation}

BendTheBow must detect whether the posterior has off-diagonal correlations requiring full Hessian computation.

\subsubsection{Perpendicular Step Fraction}

Analyze L-BFGS trajectories from GreenDragon's TrajectoryBank. In whitened coordinates (scaled by $\sqrt{-H_{ii}}$), compute for each optimization step:
\begin{equation}
    f_\perp = \frac{\|\Delta\theta_\perp\|}{\|\Delta\theta\|}
\end{equation}
where $\Delta\theta_\perp$ is the component perpendicular to the gradient direction. For axis-aligned posteriors, optimization proceeds along coordinate axes ($f_\perp \approx 0$). Significant perpendicular motion ($f_\perp > \epsilon_{\text{rot}}$) indicates rotation.

\subsubsection{Finite-Difference Probe}

If trajectory analysis is inconclusive, probe off-diagonal structure directly with 2 evaluations:
\begin{equation}
    b = \frac{v^T H v - a}{d-1}, \quad v = \frac{1}{\sqrt{d}}[1, 1, \ldots, 1]^T
\end{equation}
where $a = \text{mean}(H_{ii})$. Significant $|b|$ indicates uniform correlation requiring full Hessian.

\subsection{The $L_\infty$ Metric}
\label{app:linf}

The $L_\infty$ (Chebyshev) distance:
\begin{equation}
    \|x - y\|_\infty = \max_{i=1,\ldots,d} |x_i - y_i|
\end{equation}

\paragraph{Dimension Independence.} Unlike $L_2$ distance which diverges as $\sqrt{d}$, $L_\infty$ distance between uniform random points in $[0,1]^d$ converges to 1:
\begin{equation}
    \mathbb{E}[\|x-y\|_\infty] \approx \frac{d}{d+1} \xrightarrow{d \to \infty} 1
\end{equation}
This makes $L_\infty$ natural for high-dimensional deduplication in ChiSao.

\paragraph{Hypercube Balls.} The $L_\infty$ ball $\{y : \|x-y\|_\infty \leq r\}$ is a hypercube of side $2r$, enabling efficient spatial hashing.

\subsection{Coordinate Whitening}
\label{app:whitening}

Near a peak $\theta^*$ with Hessian $H$, the likelihood is approximately:
\begin{equation}
    \ell(\theta) \approx \ell(\theta^*) - \frac{1}{2}(\theta - \theta^*)^T H (\theta - \theta^*)
\end{equation}

The \emph{whitening transformation} $\phi = L(\theta - \theta^*)$ where $H = L^T L$ (Cholesky) gives:
\begin{equation}
    \ell(\theta) \approx \ell(\theta^*) - \frac{1}{2} \|\phi\|^2
\end{equation}

In whitened coordinates, iso-likelihood contours are spherical. The Jacobian appears in the evidence:
\begin{equation}
    \log Z_k = \ell(\theta^*_k) + \frac{d}{2}\log(2\pi) - \log|L_k|
\end{equation}

\subsection{Characteristic Scale Estimation}
\label{app:scales}

SingleWhip estimates characteristic scales from ray samples via curvature analysis.

\paragraph{Scale-Dependent Curvature.} At skip-length $h$:
\begin{equation}
    \kappa(h) = \max_t \left| \frac{f(t+h) - 2f(t) + f(t-h)}{h^2} \right|
\end{equation}

For functions with multi-scale structure $f = f_{\text{global}} + f_{\text{local}}$:
\begin{itemize}
    \item $h \ll \lambda_{\text{local}}$: $\kappa(h)$ captures local oscillations
    \item $h \gg \lambda_{\text{local}}$: $\kappa(h)$ sees only global trend
\end{itemize}

\paragraph{Scale Detection.} Compute $\kappa(h)$ at log-spaced $h$ values. Transitions where $\frac{d \log \kappa}{d \log h}$ is most negative indicate characteristic scales $\lambda_{\text{fine}}$, $\lambda_{\text{mid}}$, $\lambda_{\text{coarse}}$.

\subsection{Log-Sum-Exp Stability}
\label{app:logsumexp}

Computing $\log \sum_k e^{x_k}$ directly causes overflow/underflow. The stable form:
\begin{equation}
    \mathrm{logsumexp}(x_1, \ldots, x_K) = x_{\max} + \log \sum_{k=1}^{K} e^{x_k - x_{\max}}
\end{equation}
where $x_{\max} = \max_k x_k$. All exponents are now $\leq 0$, preventing overflow.

For multimodal evidence:
\begin{equation}
    \log Z = \mathrm{logsumexp}(\log Z_1, \ldots, \log Z_K)
\end{equation}


\end{document}